\documentclass[journal,twocolumn,10pt]{IEEEtran}

\usepackage[T1]{fontenc}
\usepackage{cite}
\usepackage[cmex10]{amsmath}
\usepackage{amsthm}
\usepackage{amssymb}
\usepackage{mathtools}

\usepackage{algorithmic}
\usepackage{textcomp}
\usepackage[usenames, dvipsnames]{color}
\usepackage[linesnumbered,ruled]{algorithm2e}
\usepackage{subcaption}
\usepackage{graphicx}  
\usepackage{url}
\usepackage{xcolor,colortbl}
\usepackage[usestackEOL]{stackengine}
\usepackage {tikz}
\usetikzlibrary {positioning}
\usepackage{multirow}
\usepackage{makecell}

\definecolor{LayerColor}{RGB}{230, 230, 230}
\definecolor{InOutColor}{RGB}{240, 243, 255}
\definecolor{cellColor}{RGB}{230, 230, 230}

\newtheorem{definition}{Definition}

\newtheorem{proposition}{Proposition}

\newtheorem{example}{Example}

\newcommand{\mc}[1]{\mathcal{#1}}

\newcommand{\mb}[1]{\mathbf{#1}}

\DeclareMathOperator*{\argmin}{arg\;min}
\DeclareMathOperator*{\argmax}{arg\;max}

\ifCLASSINFOpdf
\else
\fi

\begin{document}
%
\title{Linear and Deep Neural Network-based Receivers for Massive MIMO Systems with One-Bit ADCs}
%
%
%

\author{Ly~V.~Nguyen,  A.~Lee~Swindlehurst, and Duy~H.~N.~Nguyen
	\thanks{Ly V. Nguyen is with the Computational Science Research Center, San Diego State University, San Diego, CA, USA 92182 (e-mail: vnguyen6@sdsu.edu).}
	\thanks{A. Lee Swindlehurst is with the Center for Pervasive Communications and Computing, Samueli School of Engineering, University of
		California, Irvine, CA, USA 92697 (e-mail: swindle@uci.edu).}
	\thanks{Duy H. N. Nguyen is with the Department of Electrical and Computer Engineering, San Diego State University, San Diego, CA, USA 92182 (e-mail: duy.nguyen@sdsu.edu).}}

\maketitle

\begin{abstract}
The use of one-bit analog-to-digital converters (ADCs) is a practical solution for reducing cost and power consumption in massive Multiple-Input-Multiple-Output (MIMO) systems. However, the distortion caused by one-bit ADCs makes the data detection task much more challenging. In this paper, we propose a two-stage detection method for massive MIMO systems with one-bit ADCs. In the first stage, we propose several linear receivers based on the Bussgang decomposition that show significant performance gains over existing linear receivers. Next, we reformulate the maximum-likelihood (ML) detection problem to address its non-robustness. Based on the reformulated ML detection problem, we propose a model-driven deep neural network-based detector, namely OBMNet, whose performance is comparable with an existing support vector machine-based receiver, albeit with a much lower computational complexity. A nearest-neighbor search method is then proposed for the second stage to refine the first stage solution. Unlike existing search methods that typically perform the search over a large candidate set, the proposed search method generates a limited number of most likely candidates and thus limits the search complexity. Numerical results confirm the low complexity, efficiency, and robustness of the proposed two-stage detection method.
\end{abstract}


%
\IEEEpeerreviewmaketitle

\section{Introduction}
Massive multiple-input multiple-output (MIMO) systems, possessing the capability of boosting the throughput and energy efficiency by several orders of magnitude over conventional MIMO systems~\cite{ngo2013energy,Hoydis2013massive}, are considered to be a disruptive solution for 5G-and-beyond networks \cite{Boccardi2014Five,andrews2014will}. However, a massive MIMO system requires a large number of radio-frequency (RF) chains, which significantly increases the power consumption and hardware complexity. Among the components of an RF chain,  high-resolution analog-to-digital converters (ADCs) are power-hungry devices whose power consumption increases exponentially with the number of bits per sample and linearly with the sampling rate \cite{walden1999analog}. A promising solution for reducing the power consumption and hardware complexity is to use low-resolution ADCs. The simplest architecture involving one-bit ADCs requires only one comparator and does not require an automatic gain control (AGC). Therefore, the use of one-bit ADCs can significantly reduce both the power consumption and hardware complexity. However, the severe nonlinearity of one-bit ADCs causes significant distortions in the received signals, since only the \textit{sign} of the real and imaginary parts of the received signals is retained.

Due to the severe nonlinearity, data detection in one-bit massive MIMO systems becomes much more challenging. Numerous efforts have been made to address this problem, e.g.,~\cite{choi2016near,Jeon2018One,wen2016bayes,jeon2019robust,Song2019CRC-Aided,Cho2019OneBitSCSO,Shao2018Iterative}. A one-bit maximum-likelihood (ML) detector was derived in~\cite{choi2016near}. For large-scale systems where ML detection is impractical, the authors of~\cite{choi2016near} proposed a so-called near-ML (nML) data detection method. The ML and nML methods are however non-robust at high signal-to-noise ratios (SNRs) when the channel state information (CSI) is not perfectly known. A one-bit sphere decoding (OSD) technique was proposed in~\cite{Jeon2018One}. However, the OSD technique requires a preprocessing stage whose computational complexity is exponentially proportional to both the number of receive and transmit antennas. The exponential computational complexity of OSD makes it difficult to implement in large-scale MIMO systems. Generalized approximate message passing (GAMP) and Bayes inference are exploited in~\cite{wen2016bayes}, but the resulting method is sophisticated and expensive to implement. Several other data detection approaches have also been proposed in~\cite{jeon2019robust,Song2019CRC-Aided,Cho2019OneBitSCSO,Shao2018Iterative}, but they are only applicable in systems where either a cyclic redundancy check (CRC)~\cite{jeon2019robust,Song2019CRC-Aided,Cho2019OneBitSCSO} or an error correcting code such as a low-density parity-check (LDPC) code~\cite{Shao2018Iterative} is available. In this paper, we propose a two-stage detection method for massive MIMO systems with one-bit ADCs. The proposed method is efficient and robust with low complexity, and also applicable to large-scale systems without the need for CRC or error correcting~codes. 

In the first stage, we focus on a class of linear receivers. Existing work in this class has taken one of the following two strategies:~(i)~using standard linear receivers designed for systems with infinite-resolution ADCs, e.g.,~\cite{choi2016near,li2017channel,Jacobsson2017Throughput}; or~(ii)~using an approximate model for the one-bit ADC to construct other linear receiver designs, e.g.,~\cite{Liu2018Asymptotic,mezghani2007modified}. Here, we exploit the Bussgang decomposition~\cite{bussgang1952crosscorrelation} to propose new Bussgang-based linear receivers. Then, we study a deep learning-based detector for one-bit massive MIMO systems. There has been considerable recent interest in learning-based methods for MIMO data detection~\cite{Samuel2019Learning,Khani2020Adaptive,Nguyen2020Tabu,Gao2018Sparsely,nguyen2019supervised,Jeon2018supervised,Kim2019SemiSupervised,Nguyen2020SVM}. While the deep learning-based detectors in~\cite{Samuel2019Learning,Khani2020Adaptive,Nguyen2020Tabu,Gao2018Sparsely} are designed for MIMO systems with full-resolution ADCs, the learning-based detectors in~\cite{nguyen2019supervised,Jeon2018supervised,Kim2019SemiSupervised} are dedicated to systems with low-resolution ADCs and are ``blind'' in the sense that channel state information (CSI) is not required. However, these blind detection methods are restricted to MIMO systems with a small number of transmit antennas and only low-dimensional constellations. More recently, in~\cite{Nguyen2020SVM} a support vector machine (SVM) was exploited for one-bit MIMO data detection, and the SVM approach was shown to achieve better performance than the above linear and learning-based receivers. In this paper, we develop new linear receiver designs, as well as a new deep neural network (DNN)-based architecture, namely OBMNet, that can be implemented for one-bit massive MIMO data detection.

The contributions of the proposed receivers for this first stage are summarized as follows:
\begin{itemize}
	\item First, we exploit the Bussgang decomposition to mitigate the severe nonlinearity of one-bit ADCs and achieve a linear input-output relationship, which is then used to derive Bussgang-based linear receivers. Numerical results show that the high-SNR bit-error-rate (BER) floor of our proposed Bussgang-based linear receivers is significantly lower than that of existing methods.
	\item Next, we reformulate the ML detection problem by approximating the cumulative distribution function of a Gaussian random variable with a Sigmoid function. We show that the reformulated problem addresses the non-robustness issue of conventional ML detection. We then propose a model-driven OBMNet for data detection in one-bit massive MIMO systems. Unlike the structure of conventional DNNs where each layer contains a fixed weight matrix and a fixed bias vector, each layer of the proposed OBMNet has two adaptive weight matrices and no bias vector. Numerical results show that OBMNet outperforms the linear receivers and its performance is also comparable with that of the SVM-based method in~\cite{Nguyen2020SVM}. However, the proposed OBMNet has much lower computational complexity than the SVM-based method.
\end{itemize}

In the second stage, we propose a nearest-neighbor (NN) search method to refine the solution of stage~1. The idea of using two-stage detection methods has been studied previously in~\cite{choi2016near,Nguyen2020SVM}. However, the search metric used by the second stage of~\cite{choi2016near} is susceptible to CSI errors. This issue was addressed in~\cite{Nguyen2020SVM} thanks to a more robust search metric. Although the second stage in~\cite{Nguyen2020SVM} is robust, its complexity can be very high since the dimension of the search space over the entire candidate set can be very large. The contribution of the proposed NN search method is that it generates searches over a limited number of candidates that are nearest to the solution of stage 1 and thus helps contain the search complexity. The main challenge is to obtain the set of nearest candidates efficiently and quickly. To overcome this challenge, we propose a recursive strategy that can obtain this candidate set quickly so that the NN search method can be implemented in an efficient manner.

The rest of this paper is organized as follows: Section~\ref{sec_linear_receivers} introduces the assumed system model and presents the conventional as well as the proposed Bussgang-based linear receivers. The reformulated robust ML detection problem and OBMNet are proposed in Section~\ref{sec_robust_ML_and_DNN_receivers}. Section~\ref{sec_nearest_neighbors_search} presents the proposed NN search method. A computational complexity analysis and numerical results are given in Section~\ref{sec_complexity_analysis_and_numerical_results} and Section~\ref{sec_conclusion} concludes the paper.

\textit{Notation}: Upper-case and lower-case boldface letters denote matrices and column vectors, respectively. $\mathbb{E}[\cdot]$ represents expectation. The operator $|\cdot|$ denotes the absolute value of a number. $\|\cdot\|$ denotes the $\ell_2$-norm of a vector. The transpose and conjugate transpose are denoted by $[\cdot]^T$ and $[\cdot]^H$, respectively. The notation $\Re\{\cdot\}$ and $\Im\{\cdot\}$ respectively denotes the real and imaginary parts of the complex argument. $\mathbb{R}$ and $\mathbb{C}$ denote the set of real and complex numbers, respectively, and $j$ is the unit imaginary number satisfying $j^2=-1$. $\mc{CN}(0,\sigma^2)$ denotes a zero-mean circularly symmetric Gaussian random variable with variance $\sigma^2$, $\Phi(t) = \int_{-\infty}^{t}\frac{1}{\sqrt{2\pi}}e^{-\frac{\tau^2}{2}}d\tau$ is the cumulative distribution function of the standard Gaussian random variable and $\sigma(t) = 1/(1+e^{-t})$ is the Sigmoid function. If $\Re\{\cdot\}$, $\Im\{\cdot\}$, $\Phi(\cdot)$, and $\sigma(\cdot)$ are applied to a matrix or vector, they are applied separately to every element of that matrix or vector.

\section{Linear Receivers for First-Stage Detection}
\label{sec_linear_receivers}
This section introduces different types of linear receivers for massive MIMO systems with one-bit ADCs. We first present conventional linear receivers and then use the Bussgang decomposition to propose three new ones including Bussgang-based maximal ratio combining (BMRC), Bussgang-based zero-forcing (BZF), and Bussgang-based minimum mean squared error (BMMSE).

\subsection{System Model}
\label{subsec_system_model}
\begin{figure}[t!]
	\centering
	\includegraphics[width=\linewidth]{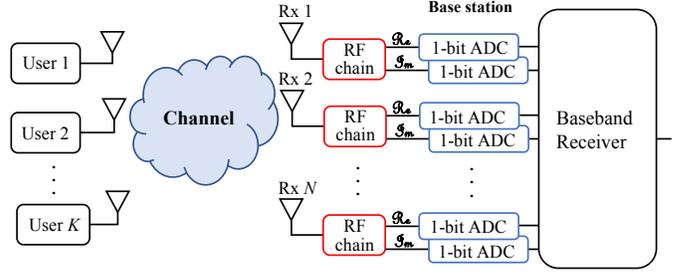}
	\caption{Block diagram of a massive MIMO system with $K$ single-antenna users and an $N$-antenna base station equipped with $2N$ one-bit ADCs.}
	\label{fig_system_model}
\end{figure}
We consider an uplink massive MIMO system as illustrated in Fig.~\ref{fig_system_model} with $K$ single-antenna users and an $N$-antenna base station, where it is assumed that $N \geq K$. Let $\bar{\mathbf{x}} = [\bar{x}_1, \bar{x}_2, \ldots, \bar{x}_K]^T \in \mathbb{C}^K$ denote the transmitted signal vector, where $\bar{x}_k$ is the signal transmitted from the $k^{\text{th}}$ user under the power constraint $\mathbb{E}[|\bar{x}_k|^2]=1$. The signal $\bar{x}_k$ is drawn from a constellation $\bar{\mathcal{M}}$, e.g, QPSK or $16$-QAM. Let $\bar{\mathbf{H}} \in \mathbb{C}^{N\times K}$ denote the channel, which is assumed to be block flat fading. Let $\bar{\mathbf{r}} = [\bar{r}_1, \bar{r}_2, \ldots, \bar{r}_N]^T \in \mathbb{C}^N$ be the unquantized received signal vector at the base station, which is given as 
\begin{equation}
\bar{\mathbf{r}} = \bar{\mathbf{H}}\bar{\mathbf{x}}+\bar{\mathbf{z}},
\label{eq_system_model_complex}
\end{equation}
where $\bar{\mathbf{z}} = [\bar{z}_1, \bar{z}_2, \ldots,\bar{z}_N]^T \in \mathbb{C}^{N}$ is a noise vector whose elements are assumed to be independent and identically distributed (i.i.d.) as $\mathcal{CN}(0,N_0)$, and $N_0$ is the noise power. Each analog received signal is then quantized by a pair of one-bit ADCs. Hence, we have the received signal
\begin{equation}
\bar{\mathbf{y}} = \operatorname{sign}(\bar{\mathbf{r}}) = \operatorname{sign}\left(\Re\{\bar{\mathbf{r}}\}\right) + j\operatorname{sign}\left(\Im\{\bar{\mathbf{r}}\}\right)
\label{eq_onebit_received_signal}
\end{equation}
where $\operatorname{sign}(\cdot)$ represents the one-bit ADC with $\operatorname{sign}(a) = +1$ if $a \geq 0$ and $\operatorname{sign}(a) = -1$ if $a < 0$. The operator $\operatorname{sign}(\cdot)$ of a matrix or vector is applied separately to every element of that matrix or vector. The SNR is defined as $\rho = 1/N_0$.

Given a received signal vector $\bar{\mathbf{y}}$ and a linear receiver represented by a combining matrix $\mathbf{W} = [\mathbf{w}_1,\mathbf{w}_2,\ldots,\mathbf{w}_K]^T \in \mathbb{C}^{K\times N}$, the demultiplexing task is performed as
\begin{equation}
\acute{\mathbf{x}} = [\acute{x}_1,\acute{x}_2,\ldots,\acute{x}_K]^T = \mathbf{W}\bar{\mathbf{y}}.
\label{eq_demultiplexing}
\end{equation}
The signal $\acute{\mathbf{x}}$ is then equalized before symbol-by-symbol detection is performed. In the following, we present different structures for the combining matrix $\mathbf{W}$. The discussion in the following sections assumes that the channel $\bar{\mathbf{H}}$ is available at the base station, but in practice an estimate of the channel would be used instead.

\subsection{Conventional Linear Receivers}
A straightforward strategy to obtain linear receivers for one-bit massive MIMO systems is to simply ignore the non-linear effect of the one-bit ADCs and use the conventional linear receivers designed for massive MIMO systems with infinite-resolution ADCs, as follows:
\begin{itemize}
	\item MRC receiver $$\mathbf{W}_{\mathtt{MRC}} = \bar{\mathbf{H}}^H,$$
	\item ZF receiver 
	$$\mathbf{W}_{\mathtt{ZF}} = \big(\bar{\mathbf{H}}^H\bar{\mathbf{H}}\big)^{-1}\bar{\mathbf{H}}^H,$$
	\item MMSE receiver
	$$\mathbf{W}_{\mathtt{MMSE}} = \big(\bar{\mathbf{H}}^H\bar{\mathbf{H}} + N_0\mathbf{I}_K\big)^{-1}\bar{\mathbf{H}}^H.$$
\end{itemize}

In another strategy, the nonlinear effect of the one-bit ADCs can be linearized by the Additive Quantization Noise Model (AQNM) \cite{Fletcher2007roubust,Orhan2015low} as
\begin{equation}
\bar{\mathbf{y}} = \kappa\bar{\mathbf{r}} + \bar{\mathbf{d}} = \kappa \bar{\mathbf{H}}\bar{\mathbf{x}} + \kappa \bar{\mathbf{z}} + \bar{\mathbf{d}},
\label{eq_AQNM_model}
\end{equation}
where $\kappa = 1 - \alpha$ and $\alpha$ is the inverse of the signal-to quantization-noise ratio, which for one-bit ADCs is approximately given by $\alpha \approx 0.3634$ \cite{Orhan2015low}. The quantization distortion $\bar{\mathbf{d}}$ is treated as additive Gaussian noise $\bar{\mathbf{d}} \sim \mathcal{CN}(\mathbf{0},\boldsymbol{\Sigma}_{\bar{d}})$ that is uncorrelated with $\bar{\mathbf{r}}$, where
$
\boldsymbol{\Sigma}_{\bar{d}} = \alpha \kappa \operatorname{diag}(\bar{\mathbf{H}}\bar{\mathbf{H}}^H+N_0\mathbf{I}_N).
$
The MMSE receiver for the model in~(\ref{eq_AQNM_model}) is given as \cite{Liu2018Asymptotic}
\begin{equation}
\mathbf{W}_{\mathtt{AQNM-MMSE}} = \bar{\mathbf{H}}^H\bigg(\bar{\mathbf{H}}\bar{\mathbf{H}}^H + \frac{1}{\kappa^2}\boldsymbol{\Sigma}_{\bar{d}} + N_0\mathbf{I}_N\bigg)^{-1}.
\end{equation}
Another approximate MMSE receiver for quantized MIMO systems, referred to as  the ``Wiener Filter on Quantized data'' (WFQ), is proposed in \cite{mezghani2007modified} as
\begin{equation}
\mathbf{W}_{\mathtt{WFQ}} = \bar{\mathbf{H}}^H\Big(\kappa \mathbf{\Sigma}_{\bar{r}} + \alpha \operatorname{diag}(\mathbf{\Sigma}_{\bar{r}})\Big)^{-1},
\end{equation}
where $\mathbf{\Sigma}_{\bar{r}} = \bar{\mathbf{H}}\bar{\mathbf{H}}^H + N_0\mathbf{I}_N$ is the covariance matrix of $\bar{\mathbf{r}}$.

Once a combining matrix $\mathbf{W}$ has been computed, the demultiplexing task can be performed as in (\ref{eq_demultiplexing}). If the combining matrix is $\mathbf{W}_{\mathtt{MRC}}$, then the signal $\acute{\mathbf{x}}$ is equalized as
\begin{equation}
\check{x}_k = \frac{\acute{x}_k}{\mathbf{w}_k^T\bar{\mathbf{h}}_k},
\end{equation}
where $\mathbf{w}_k$ is the $k^{\text{th}}$ column of $\mathbf{W}_{\mathtt{MRC}}$. Since the norm squared of $\check{\mathbf{x}} = [\check{x}_1,\check{x}_2,\ldots,\check{x}_K]^T$ may not equal $K$, the signal $\check{\mathbf{x}}$ should be rescaled as \cite{choi2016near}
\begin{equation}
\tilde{\mathbf{x}}= [\tilde{x}_1, \tilde{x}_2,\ldots,\tilde{x}_K]^T = \sqrt{K}\frac{\check{\mathbf{x}}}{\|\check{\mathbf{x}}\|_2}.
\label{eq_rescaled_signal}
\end{equation}
Finally, the signal $\dot{\mathbf{x}}$ can be used for symbol-by-symbol detection:
\begin{equation}
\hat{x}_k = \argmax_{\bar{x}\in\bar{\mathcal{M}}} |\bar{x}-\tilde{x}_k|.
\label{eq_sym_by_sym_detection}
\end{equation}

\subsection{Proposed Bussgang-Based Linear Receivers}
Here, we exploit the Bussgang decomposition to linearize the system model $\bar{\mathbf{y}} = \operatorname{sign}(\bar{\mathbf{r}})$ and then use the linearized model to propose new MRC, ZF, and MMSE receiver structures. Following the Bussgang decomposition, the system model $\bar{\mathbf{y}} = \operatorname{sign}(\bar{\mathbf{r}})$ can be rewritten as $\bar{\mathbf{y}} = \bar{\mathbf{V}}\bar{\mathbf{r}} + \bar{\mathbf{e}}$ \cite{mezghani2012capacity} where $\bar{\mathbf{e}}$ is the quantization distortion, which is uncorrelated with $\bar{\mathbf{r}}$, i.e., $\mathbb{E}\big[\bar{\mathbf{r}}\bar{\mathbf{e}}^H\big] = \mathbb{E}\big[\bar{\mathbf{r}}\big]\mathbb{E}\big[\bar{\mathbf{e}}^H\big]$, and 
\begin{equation}
\bar{\mathbf{V}} = \sqrt{\frac{2}{\pi}}\operatorname{diag}(\boldsymbol{\Sigma}_{\bar{r}})^{-\frac{1}{2}}.
\end{equation} 
Let $\bar{\mathbf{A}} = \bar{\mathbf{V}}\bar{\mathbf{H}}$ and $\bar{\mathbf{n}} = \bar{\mathbf{V}}\bar{\mathbf{z}} + \bar{\mathbf{e}}$, so the system model becomes
\begin{equation}
\bar{\mathbf{y}} = \bar{\mathbf{A}}\bar{\mathbf{x}} + \bar{\mathbf{n}},
\label{eq_Bussgang_based_model}
\end{equation}
where $\bar{\mathbf{A}} = \sqrt{2/\pi}\operatorname{diag}(\boldsymbol{\Sigma}_{\bar{r}})^{-\frac{1}{2}}\bar{\mathbf{H}}$ is the effective channel and $\bar{\mathbf{n}}$ is the effective noise, which is modeled as Gaussian with zero mean and covariance matrix  \cite{mezghani2012capacity}:
\begin{equation}\label{Sigma-n}
\begin{split}
\boldsymbol{\Sigma}_{\bar{n}} =& \frac{2}{\pi}\Big[\operatorname{arcsin}\Big(\operatorname{diag}(\mathbf{\Sigma}_{\bar{r}})^{-\frac{1}{2}}\mathbf{\Sigma}_{\bar{r}}\operatorname{diag}(\mathbf{\Sigma}_{\bar{r}})^{-\frac{1}{2}}\Big)-\\
&\quad\;\operatorname{diag}(\mathbf{\Sigma}_{\bar{r}})^{-\frac{1}{2}}\mathbf{\Sigma}_{\bar{r}}\operatorname{diag}(\mathbf{\Sigma}_{\bar{r}})^{-\frac{1}{2}}+ N_0\operatorname{diag}(\mathbf{\Sigma}_{\bar{r}})^{-1}\Big].
\end{split}
\end{equation}
Note that $\operatorname{arcsin}(\mathbf{C}) = \operatorname{arcsin}(\Re\{\mathbf{C}\}) + j\operatorname{arcsin}(\Im \{\mathbf{C}\})$ for any complex matrix $\mathbf{C}$, and the operation $\operatorname{arcsin}(\mathbf{\cdot})$ of a real matrix is applied separately on each element of that matrix. 

Based on the effective channel $\bar{\mathbf{A}}$, we can derive a Bussgang-based MRC (BMRC) receiver and a Bussgang-based ZF (BZF) receiver as
\begin{equation}
\mathbf{W}_{\mathtt{BMRC}} = \bar{\mathbf{A}}^H = \sqrt{\frac{2}{\pi}}\bar{\mathbf{H}}^H\operatorname{diag}(\boldsymbol{\Sigma}_{\bar{r}})^{-\frac{1}{2}},
\end{equation}
and
\begin{align}
\mathbf{W}_{\mathtt{BZF}} &= (\bar{\mathbf{A}}^H\bar{\mathbf{A}})^{-1}\bar{\mathbf{A}}^H \notag\\
&=\sqrt{\frac{\pi}{2}} \big(\bar{\mathbf{H}}^H\operatorname{diag}(\boldsymbol{\Sigma}_{\bar{r}})^{-1}\bar{\mathbf{H}}\big)^{-1}\bar{\mathbf{H}}^H\operatorname{diag}(\boldsymbol{\Sigma}_{\bar{r}})^{-\frac{1}{2}}.
\end{align}

We now derive the MMSE receiver for this Bussgang-based system model. The Bussgang-based MMSE (BMMSE) receiver can be obtained by solving the following optimization problem:
\begin{equation}
\begin{aligned}
& \underset{\{\mathbf{W}\}}{\operatorname{minimize}} 
& & \mathbb{E}\big[\|\bar{\mathbf{x}} - \mathbf{W}\bar{\mathbf{y}}\|_2^2\big],
\end{aligned}
\label{eq_BMMSE_problem}
\end{equation}
whose solution is given in closed form as follows:
\begin{eqnarray}
\mb{W}_{\mathtt{BMMSE}} = \mathbb{E}\big[\bar{\mathbf{x}}\bar{\mathbf{y}}^H\big]
\big(\mathbb{E}\big[\bar{\mathbf{y}}\bar{\mathbf{y}}^H\big]\big)^{-1}.
\end{eqnarray}
We can expand $\mathbb{E}\big[\bar{\mathbf{x}}\bar{\mathbf{y}}^H\big] = \mathbb{E}\left[\bar{\mb{x}}\bar{\mb{x}}^H\bar{\mb{A}}\right] + \mathbb{E}\left[\bar{\mb{x}}\mb{n}^H\right] = \bar{\mb{A}} $
due to $\mathbb{E}\big[\bar{\mathbf{x}}\bar{\mathbf{x}}^H\big] = \mathbf{I}_K$ and
$\mathbb{E}\left[\bar{\mb{x}}\mb{n}^H\right]  = \mb{0}$. We have $\mathbb{E}\big[\bar{\mathbf{x}}\bar{\mathbf{n}}^H\big] = \mathbf{0}$ since
\begin{equation*}
\mathbb{E}\big[\bar{\mathbf{x}}\bar{\mathbf{n}}^H\big] = \mathbb{E}\big[\bar{\mathbf{x}}(\bar{\mathbf{V}}\bar{\mathbf{z}} + \bar{\mathbf{e}})^H\big]
= \mathbb{E}\big[\bar{\mathbf{x}}\bar{\mathbf{z}}^H\big]\bar{\mathbf{V}} + \mathbb{E}\big[\bar{\mathbf{x}}\bar{\mathbf{e}}^H\big],
\end{equation*}
where $\mathbb{E}\big[\bar{\mathbf{x}}\bar{\mathbf{z}}^H\big] = \mathbb{E}\big[\bar{\mathbf{x}}\big] \mathbb{E}\big[\bar{\mathbf{z}}^H\big] = \mathbf{0}$, and $\mathbb{E}\big[\bar{\mathbf{x}}\bar{\mathbf{e}}^H\big] = \mathbf{0}$ since
\begin{equation*}
\begin{cases}
\mathbb{E}\big[\bar{\mathbf{r}}\bar{\mathbf{e}}^H\big] = \bar{\mathbf{H}}\mathbb{E}\big[\bar{\mathbf{x}}\bar{\mathbf{e}}^H\big] + \mathbb{E}\big[\bar{\mathbf{z}}\bar{\mathbf{e}}^H\big],\\
\mathbb{E}\big[\bar{\mathbf{r}}\bar{\mathbf{e}}^H\big] = \mathbb{E}\big[\bar{\mathbf{r}}\big]\mathbb{E}\big[\bar{\mathbf{e}}^H\big] = \mathbf{0},\\
\mathbb{E}\big[\bar{\mathbf{z}}\bar{\mathbf{e}}^H\big] = \mathbf{0}.
\end{cases}
\end{equation*}
In addition, $\mathbb{E}\big[\bar{\mathbf{y}}\bar{\mathbf{y}}^H\big]$ is given by \cite{mezghani2012capacity}
$$\mathbb{E}\big[\bar{\mathbf{y}}\bar{\mathbf{y}}^H\big] =  
\frac{2}{\pi}\operatorname{arcsin}\Big(\operatorname{diag}(\mathbf{\Sigma}_{\bar{r}})^{-\frac{1}{2}}\mathbf{\Sigma}_{\bar{r}}\operatorname{diag}(\mathbf{\Sigma}_{\bar{r}})^{-\frac{1}{2}}\Big).$$
Hence, the resulting BMMSE receiver is given as
\begin{align}
	\mathbf{W}_{\mathtt{BMMSE}} &= \bar{\mathbf{A}}^H\left[\frac{2}{\pi}\operatorname{arcsin}\Big(\operatorname{diag}(\mathbf{\Sigma}_{\bar{r}})^{-\frac{1}{2}}\mathbf{\Sigma}_{\bar{r}}\operatorname{diag}(\mathbf{\Sigma}_{\bar{r}})^{-\frac{1}{2}}\Big)\right]^{-1}\nonumber\\
	&= \bar{\mathbf{A}}^H\left(\bar{\mathbf{A}}\bar{\mathbf{A}}^H + \boldsymbol{\Sigma}_{\bar{n}}\right)^{-1} \; ,
\end{align}
where the second equality comes from the equivalent model in~\eqref{eq_Bussgang_based_model} and the expression for $\boldsymbol{\Sigma}_{\bar{n}}$ in \eqref{Sigma-n}.

It can be seen that the structure of the BMMSE receiver is similar to the that of the MMSE receiver, except that the BMMSE receiver applies a new effective channel and a new effective noise covariance. These differences come as the result of linearizing the system model with the Bussgang decomposition.

Since the effective channel is $\bar{\mathbf{A}}$, if the BMRC receiver is used, the equalization step is now performed as
\begin{equation}
\check{x}_k = \frac{\acute{x}_k}{\mathbf{w}_k^T\bar{\mathbf{a}}_k},
\end{equation}
where $\mathbf{w}_k$ and $\bar{\mathbf{a}}_k$ are the $k^{\text{th}}$ column of $\mathbf{W}_{\mathtt{BMRC}}$ and $\bar{\mathbf{A}}$, respectively. The rescaling step and symbol-by-symbol detection are the same as in (\ref{eq_rescaled_signal}) and (\ref{eq_sym_by_sym_detection}).

\section{DNN-based Receiver for First-Stage Detection}
\label{sec_robust_ML_and_DNN_receivers}
In this section, we first reformulate the conventional ML rule for one-bit MIMO systems, which is then exploited to devise OBMNet. We consider the same system model as presented in Section~\ref{sec_linear_receivers}, but for convenience in later derivations, we convert~\eqref{eq_system_model_complex} and~\eqref{eq_onebit_received_signal} into the real domain as follows:
\begin{equation}
\mathbf{y} = \operatorname{sign}\left(\mathbf{H}\mathbf{x} + \mathbf{z}\right),
\label{eq_flat_fading_system_model_real_domain}
\end{equation}
where
\begin{align*}
\mathbf{y} &= \begin{bmatrix}
\Re \{\bar{\mathbf{y}}\} \\ \Im \{\bar{\mathbf{y}}\}
\end{bmatrix} \in \mathbb{R}^{2N},\ \mathbf{x} = \begin{bmatrix}
\Re \{\bar{\mathbf{x}}\} \\ \Im \{\bar{\mathbf{x}}\}
\end{bmatrix} \in \mathbb{R}^{2K},\\ 
\mathbf{z} & = \begin{bmatrix}
\Re \{\bar{\mathbf{z}}\} \\ \Im \{\bar{\mathbf{z}}\}
\end{bmatrix} \in \mathbb{R}^{2N}, \text{ and}\\
\mathbf{H} &= \begin{bmatrix}
\Re \{\bar{\mathbf{H}}\} & -\Im \{\bar{\mathbf{H}}\}\\
\Im \{\bar{\mathbf{H}}\} & \Re \{\bar{\mathbf{H}}\}
\end{bmatrix} \in \mathbb{R}^{2N\times 2K}.
\end{align*}
We also denote $\mathbf{y} = [y_1, \ldots, y_{2N}]^T$, $\mathbf{x} = [x_1, \ldots, x_{2K}]^T$, $\mathbf{z} = [z_1, \ldots, z_{2N}]^T$, and $\mathbf{H} = [\mathbf{h}_1, \ldots, \mathbf{h}_{2N}]^T$.

The conventional ML detection problem~\cite{choi2016near} for one-bit ADCs is given as
\begin{equation}
\hat{\mathbf{x}}_{\mathtt{ML}} = \argmax_{\bar{\mathbf{x}}\in \bar{\mathcal{M}}^{K}} \prod_{n=1}^{2N}\Phi (\sqrt{2\rho}y_n\hat{\mathbf{h}}_n^T\mathbf{x}),
\label{eq_conventional_ML_detection}
\end{equation}
which can also be written as
\begin{equation}
\hat{\mathbf{x}}_{\mathtt{ML}} = \argmax_{\bar{\mathbf{x}}\in \bar{\mathcal{M}}^{K}} \sum_{n=1}^{2N}\log \Phi (\sqrt{2\rho}y_n\hat{\mathbf{h}}_n^T\mathbf{x}),
\label{eq_conventional_logML_detection}
\end{equation}
where $\hat{\mathbf{h}}_n$ is an estimate of $\mathbf{h}_n$ for $n \in \{1,\ldots,2N\}$. The ML detection formulations in~\eqref{eq_conventional_ML_detection} and~\eqref{eq_conventional_logML_detection} are however non-robust at high SNRs when $\hat{\mathbf{h}}_n \neq \mathbf{h}_n$, or in other words, when the CSI is imperfectly known. This non-robustness issue is due to the function $\Phi(\cdot)$ which approaches $0$ exponentially fast and has been reported in~\cite{nguyen2019supervised,Jeon2018supervised}. A detailed explanation for this issue can be found in~\cite{Nguyen2020SVM}.
\begin{figure}[t!]
	\centering
	\begin{tikzpicture}
	\node [above] at (0.4,10) {$x^{(0)}_1$};
	\draw [semithick,->] (0,10) to (0.89,10);
	
	\node [above] at (0.4,9) {$x^{(0)}_2$};
	\draw [semithick,->] (0,9) to (0.89,9);
	
	\node [above] at (0.4,7) {$x^{(0)}_{2K}$};
	\draw [semithick,->] (0,7) to (0.89,7);
	
	\draw [fill=LayerColor, thick, rounded corners] (0.9,6.5) rectangle (1.9,10.5);
	\node at (1.4,8.5) {Layer};
	\node at (1.4,8.1) {$1$};
	
	\node [above] at (2.4,10) {$x^{(1)}_1$};
	\draw [semithick,->] (1.9,10) to (2.89,10);
	
	\node [above] at (2.4,9) {$x^{(1)}_2$};
	\draw [semithick,->] (1.9,9) to (2.89,9);
	
	\node [above] at (2.39,7) {$x^{(1)}_{2K}$};
	\draw [semithick,->] (1.9,7) to (2.89,7);
	
	\draw [fill=LayerColor, thick, rounded corners] (2.9,6.5) rectangle (3.9,10.5);
	\node at (3.4,8.5) {Layer};
	\node at (3.4,8.1) {$2$};
	
	\node [above] at (4.4,10) {$x^{(2)}_1$};
	\draw [semithick,->] (3.9,10) to (4.9,10);
	
	\node [above] at (4.4,9) {$x^{(2)}_2$};
	\draw [semithick,->] (3.9,9) to (4.9,9);
	
	\node [above] at (4.39,7) {$x^{(2)}_{2K}$};
	\draw [semithick,->] (3.9,7) to (4.9,7);
	
	\node at (5.3,10) {\textbf{\ldots}};
	\node at (5.3,9) {\textbf{\ldots}};
	\node at (5.3,7) {\textbf{\ldots}};
	
	\node [above] at (6.2,10) {$x^{(L-1)}_1$};
	\draw [semithick,->] (5.6,10) to (6.79,10);
	
	\node [above] at (6.2,9) {$x^{(L-1)}_2$};
	\draw [semithick,->] (5.6,9) to (6.79,9);
	
	\node [above] at (6.15,7) {$x^{(L-1)}_{2K}$};
	\draw [semithick,->] (5.6,7) to (6.79,7);
	
	\draw [fill=LayerColor, thick, rounded corners] (6.8,6.5) rectangle (7.8,10.5);
	\node at (7.3,8.5) {Layer};
	\node at (7.3,8.1) {$L$};
	
	\node [above] at (8.2,10) {$x^{(L)}_1$};
	\draw [semithick,->] (7.8,10) to (8.7,10);
	\node [above] at (8.2,9) {$x^{(L)}_2$};
	\draw [semithick,->] (7.8,9) to (8.7,9);
	\node [above] at (8.2,7) {$x^{(L)}_{2K}$};
	\draw [semithick,->] (7.8,7) to (8.7,7);
	
	\node at (0.4,8.25) {\textbf{\vdots}};
	\node at (2.4,8.25) {\textbf{\vdots}};
	\node at (4.4,8.25) {\textbf{\vdots}};
	\node at (6.2,8.25) {\textbf{\vdots}};
	\node at (8.3,8.25) {\textbf{\vdots}};
	\end{tikzpicture}
	\caption{Overall structure of the proposed OBMNet.}
	\label{fig_DNN-based_receiver_flat-fading}
\end{figure}
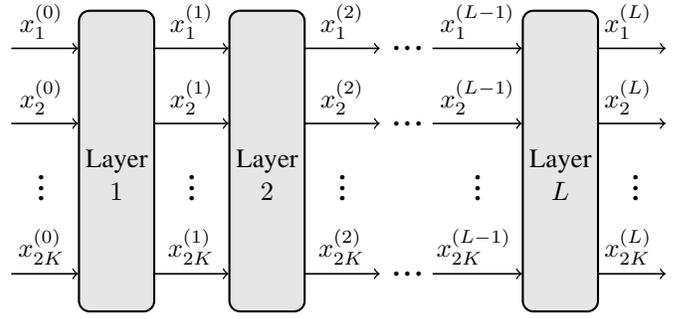
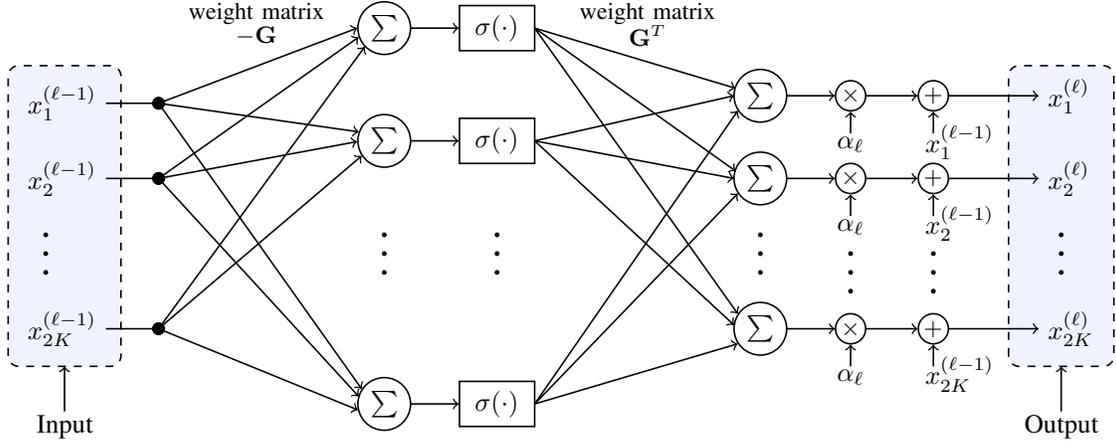
\begin{figure*}[t!]
	\centering
	\begin{tikzpicture}
	\draw [fill=InOutColor, dashed, semithick, rounded corners] (-1,6.5) rectangle (0.5,10.5);
	\draw [semithick,->] (-0.25,5.9) to (-0.25,6.5);
	\node at (-0.25,5.7) {Input};
	
	\draw [fill=InOutColor, dashed, semithick, rounded corners] (12.3,6.5) rectangle (13.7,10.5);
	\draw [semithick,->] (13,5.9) to (13,6.5);
	\node at (13,5.7) {Output};
	
	\node [left] at (0.3,10) {$x^{(\ell-1)}_1$};
	\draw [semithick] (0.3,10) to (1,10);
	\draw [fill] (1,10) circle [radius=0.08];
	
	\draw [semithick] (4,11) circle [radius=0.35];
	\node at (4,11) {$\sum$};
	\draw [semithick,->] (4.35,11) to (4.99,11);
	\draw [semithick] (5,10.7) rectangle (6,11.3);
	\node at (5.5,11) {$\sigma(\cdot)$};
	
	\draw [semithick] (9,10.1) circle [radius=0.35];
	\node at (9,10.1) {$\sum$};
	\draw [semithick,->] (6,11) to (8.67,10.2);
	\draw [semithick,->] (6,9.5) to (8.65,10.1);
	\draw [semithick,->] (6,6) to (8.7,9.9);
	
	\draw [semithick, ->] (1,10) to (3.64,11);
	\draw [semithick, ->] (1,9) to (3.66,10.88);
	\draw [semithick, ->] (1,7) to (3.74,10.75);
	
	\draw [semithick, ->] (9.35,10.1) to (9.99,10.1);
	\draw [semithick] (10.2,10.1) circle [radius=0.2];
	\node at (10.2,10.1) {\small {$\times$}};
	\draw [semithick,->] (10.2,9.6) to (10.2,9.89);
	\node at (10.2,9.45) {$\alpha_{\ell}$};
	\draw [semithick, ->] (10.4,10.1) to (11.09,10.1);
	\draw [semithick] (11.3,10.1) circle [radius=0.2];
	\draw [semithick, ->] (11.3,9.6) to (11.3,9.89);
	\node at (11.65,9.49) {$x^{(\ell-1)}_1$};
	\node at (11.3,10.1) {\small{$+$}};
	\draw [semithick, ->] (11.5,10.1) to (12.7,10.1);
	\node [right] at (12.7,10.1) {$x^{(\ell)}_1$};
	
	\node [left] at (0.3,9) {$x^{(\ell-1)}_2$};
	\draw [semithick] (0.3,9) to (1,9);
	\draw [fill] (1,9) circle [radius=0.08];
	
	\draw [semithick] (4,9.5) circle [radius=0.35];
	\node at (4,9.5) {$\sum$};
	\draw [semithick,->] (4.35,9.5) to (4.99,9.5);
	\draw [semithick] (5,9.2) rectangle (6,9.8);
	\node at (5.5,9.5) {$\sigma(\cdot)$};
	
	\draw [semithick] (9,9) circle [radius=0.35];
	\node at (9,9) {$\sum$};
	\draw [semithick,->] (6,11) to (8.67,9.1);
	\draw [semithick,->] (6,9.5) to (8.65,9);
	\draw [semithick,->] (6,6) to (8.7,8.8);
	
	\draw [semithick, ->] (1,10) to (3.66,9.62);
	\draw [semithick, ->] (1,9) to (3.64,9.49);
	\draw [semithick, ->] (1,7) to (3.7,9.3);
	
	\draw [semithick, ->] (9.35,9) to (9.99,9);
	\draw [semithick] (10.2,9) circle [radius=0.2];
	\node at (10.2,9) {\small {$\times$}};
	\draw [semithick,->] (10.2,8.5) to (10.2,8.79);
	\node at (10.2,8.35) {$\alpha_{\ell}$};
	\draw [semithick, ->] (10.4,9) to (11.09,9);
	\draw [semithick] (11.3,9) circle [radius=0.2];
	\draw [semithick, ->] (11.3,8.5) to (11.3,8.79);
	\node at (11.65,8.4) {$x^{(\ell-1)}_2$};
	\node at (11.3,9) {\small {$+$}};
	\draw [semithick, ->] (11.5,9) to (12.7,9);
	\node [right] at (12.7,9) {$x^{(\ell)}_2$};
	
	\node [left] at (0.3,7) {$x^{(\ell-1)}_{2K}$};
	\draw [semithick] (0.3,7) to (1,7);
	\draw [fill] (1,7) circle [radius=0.08];
	
	\draw [semithick] (4,6) circle [radius=0.35];
	\node at (4,6) {$\sum$};
	\draw [semithick,->] (4.35,6) to (4.99,6);
	\draw [semithick] (5,5.7) rectangle (6,6.3);
	\node at (5.5,6) {$\sigma(\cdot)$};
	
	\draw [semithick] (9,7) circle [radius=0.35];
	\node at (9,7) {$\sum$};
	\draw [semithick,->] (6,11) to (8.7,7.2);
	\draw [semithick,->] (6,9.5) to (8.65,7.05);
	\draw [semithick,->] (6,6) to (8.7,6.8);
	
	\draw [semithick, ->] (1,10) to (3.7,6.2);
	\draw [semithick, ->] (1,9) to (3.66,6.1);
	\draw [semithick, ->] (1,7) to (3.65,5.92);
	
	\draw [semithick, ->] (9.35,7) to (9.99,7);
	\draw [semithick] (10.2,7) circle [radius=0.2];
	\node at (10.2,7) {\small {$\times$}};
	\draw [semithick,->] (10.2,6.5) to (10.2,6.79);
	\node at (10.2,6.35) {$\alpha_{\ell}$};
	\draw [semithick, ->] (10.4,7) to (11.09,7);
	\draw [semithick] (11.3,7) circle [radius=0.2];
	\draw [semithick, ->] (11.3,6.5) to (11.3,6.79);
	\node at (11.65,6.4) {$x^{(\ell-1)}_{2K}$};
	\node at (11.3,7) {\small {$+$}};
	\draw [semithick, ->] (11.5,7) to (12.7,7);
	\node [right] at (12.7,7) {$x^{(\ell)}_{2K}$};
	
	\draw [fill] (-0.5,8.25) circle [radius=0.025];
	\draw [fill] (-0.5,8) circle [radius=0.025];
	\draw [fill] (-0.5,7.75) circle [radius=0.025];
	
	\draw [fill] (4,8.25) circle [radius=0.025];
	\draw [fill] (4,8) circle [radius=0.025];
	\draw [fill] (4,7.75) circle [radius=0.025];
	
	\draw [fill] (5.5,8.25) circle [radius=0.025];
	\draw [fill] (5.5,8) circle [radius=0.025];
	\draw [fill] (5.5,7.75) circle [radius=0.025];
	
	\draw [fill] (9,8.25) circle [radius=0.025];
	\draw [fill] (9,8) circle [radius=0.025];
	\draw [fill] (9,7.75) circle [radius=0.025];
	
	\draw [fill] (10.2,8) circle [radius=0.025];
	\draw [fill] (10.2,7.75) circle [radius=0.025];
	\draw [fill] (10.2,7.5) circle [radius=0.025];
	
	\draw [fill] (11.3,8) circle [radius=0.025];
	\draw [fill] (11.3,7.75) circle [radius=0.025];
	\draw [fill] (11.3,7.5) circle [radius=0.025];
	
	\draw [fill] (13,8.25) circle [radius=0.025];
	\draw [fill] (13,8) circle [radius=0.025];
	\draw [fill] (13,7.75) circle [radius=0.025];
	
	\node at (2.3,11.2) {\small {weight matrix}};
	\node at (2.3,10.9) {\small {$-\mathbf{G}$}};
	
	\node at (7.5,11.2) {\small {weight matrix}};
	\node at (7.5,10.9) {\small {$\mathbf{G}^T$}};
	\end{tikzpicture}
	\caption{Specific structure of layer $\ell$ where the trainable parameter is $\alpha_{\ell}$ and the weight matrices are adaptive to the channel and the received signal.}
	\label{fig_structure_layer_l}
\end{figure*}

To address the non-robustness of the above ML formulation, we exploit a result in~\cite{bowling2009logistic}, which shows that the function $\Phi(t)$ can be accurately approximated by the Sigmoid function $\sigma(t)$, which is a widely-used activation function in machine learning research. The approximation of $\Phi(t)$ is given as
\begin{equation}
	\Phi(t) \approx \sigma(ct) = \frac{1}{1+e^{-ct}},
	\label{eq_approximate_Phi_as_Sigma}
\end{equation}
where $c = 1.702$ is a constant. It was shown in~\cite{bowling2009logistic} that $|\Phi(t)-\sigma(ct)|\leq 0.0095$, $\forall t\in \mathbb{R}$. Thus, maximizing $\log\Phi(t)$ is approximately equivalent to minimizing $\log(1+e^{-ct})$.

Applying the approximation in~\eqref{eq_approximate_Phi_as_Sigma} to~\eqref{eq_conventional_logML_detection}, we obtain the following ML detection problem:
\begin{equation}
\hat{\mathbf{x}}^{\mathrm{robust}}_{\mathtt{ML}} = \argmin_{\bar{\mathbf{x}}\in \bar{\mathcal{M}}^{K}} \sum_{n=1}^{2N}\log \left(1+e^{-c\sqrt{2\rho}y_n\hat{\mathbf{h}}_n^T\mathbf{x}}\right).
\label{eq_robust_ML_detection}
\end{equation}
As mentioned earlier, the ML detection formulation in~\eqref{eq_conventional_ML_detection} and~\eqref{eq_conventional_logML_detection} is not robust against imperfect CSI due to the $\Phi(\cdot)$ function. However, the reformulated ML detection problem~\eqref{eq_robust_ML_detection} does not share this robustness problem. It is interesting to note that $\log(1+e^t)$ is referred to as the SoftPlus activation function in the machine learning literature. Hence, the proposed robust ML detection problem in~\eqref{eq_robust_ML_detection} can be interpreted as a minimization problem whose objective is a sum of SoftPlus activation functions.

Now, we develop a DNN-based receiver based on the proposed robust ML detection problem in~\eqref{eq_robust_ML_detection}. We relax the constraint $\bar{\mathbf{x}}\in \bar{\mathcal{M}}^{K}$ in~\eqref{eq_robust_ML_detection} to $\bar{\mathbf{x}}\in \mathbb{C}^{K}$ and denote the channel estimate $\hat{\mathbf{H}} = [\hat{\mathbf{h}}_1,\ldots,\hat{\mathbf{h}}_{2N}]^T$. Let $\mathbf{G} = \operatorname{diag}(y_1,\ldots,y_{2N})\hat{\mathbf{H}}$ and define the rows of $\mathbf{G}$ as $\mathbf{G} = [\mathbf{g}_1, \ldots, \mathbf{g}_{2N}]^T$. Then~\eqref{eq_robust_ML_detection} can be rewritten as
\begin{equation}
\argmin_{\bar{\mathbf{x}}\in \mathbb{C}^{K}} \underbrace{\sum_{n=1}^{2N}\log \left(1+e^{-c\sqrt{2\rho}\mathbf{g}_n^T\mathbf{x}}\right)}_{\mathcal{P}(\mathbf{x})}.
\label{eq_relaxed_robust_ML_detection}
\end{equation}
The gradient of $\mathcal{P}(\mathbf{x})$ is
\begin{align}
\nabla \mathcal{P}(\mathbf{x}) & = \sum_{n=1}^{2N}\frac{-c\sqrt{2\rho}\,\mathbf{g}_n}{1+e^{c\sqrt{2\rho}\,\mathbf{g}_n^T\mathbf{x}}} \nonumber\\
& = -c\sqrt{2\rho}\mathbf{G}^T \sigma \big(-c\sqrt{2\rho}\mathbf{G}\mathbf{x}\big).
\end{align}
Hence, an iterative gradient descent method can be used to solve~\eqref{eq_relaxed_robust_ML_detection} as follows:
\begin{equation}
\mathbf{x}^{(\ell)} = \mathbf{x}^{(\ell-1)} + \alpha_{\ell} c\sqrt{2\rho}\, \mathbf{G}^T \sigma \left(-c\sqrt{2\rho}\mathbf{G}\mathbf{x}^{(\ell-1)}\right)
\label{eq_iterative_method}
\end{equation}
where $\ell$ is the iteration index and $\alpha_{\ell}$ is the step size. 

In order to optimize the step sizes $\{\alpha_{\ell}\}$, we use the \emph{deep unfolding} technique \cite{Hershey-Unfolding-2014} to unfold each iteration in~\eqref{eq_iterative_method} as a layer of a deep neural network. The overall structure of the proposed OBMNet is illustrated in Fig.~\ref{fig_DNN-based_receiver_flat-fading}, where there are $L$ layers and each layer takes a vector of $2K$ elements as the input and generates an output vector of the same size. The specific structure for each layer $\ell$  is illustrated in Fig.~\ref{fig_structure_layer_l}. 

It can be seen that the proposed layer structure in Fig.~\ref{fig_structure_layer_l} is different from that of conventional DNNs, since it exploits the specific structure of the ML detection problem. In particular, each layer of a conventional DNN often contains a weight matrix and a bias vector to be trained. However, due to the structure of the ML detection problem, in each layer of OBMNet the only trainable parameter is the step size $\alpha_{\ell}$. The proposed layer structure has two weight matrices $-\mathbf{G}$ and $\mathbf{G}^T$ and no bias vector, and the weight matrices are defined by the channel estimate and the received signal. 

Since $\mathbf{G} \in \mathbb{R}^{2N\times2K}$, the learning process of each layer can be interpreted as first up-converting the signal from dimension $2K$ to dimension $2N$ using the weight matrix $-\mathbf{G}$, then applying nonlinear activation functions before down-converting the signal back to dimension $2K$ using the weight matrix $\mathbf{G}^T$. The activation function in OBMNet is the Sigmoid function, which is also widely used in conventional DNNs. Note that the use of the Sigmoid activation function in OBMNet is not arbitrary but results from the use of the approximation in~\eqref{eq_approximate_Phi_as_Sigma} and the structure of the ML detection problem.

The objective function to be minimized during the training phase is $\|\tilde{\mathbf{x}} - \mathbf{x}\|^2,$
where 
\begin{equation}
	\tilde{\mathbf{x}}=\frac{\sqrt{K}}{\|\mathbf{x}^{(L)}\|}\mathbf{x}^{(L)}
	\label{eq_normalized_DNN_output}
\end{equation}
and $\mathbf{x}$ is the target signal, i.e., the transmitted signal. It should also be noted that the layered structure in Fig.~\ref{fig_structure_layer_l} does not contain the coefficient $c\sqrt{2\rho}$. We omit this coefficient because it is a constant throughout the layers of OBMNet, and the output of the last layer $\mathbf{x}^{(L)}$ needs to be normalized as in~\eqref{eq_normalized_DNN_output}. We found by experiments that this omission not only helps improve the detection performance but also helps the training process to stably converge. 

The training process is accomplished offline. A training sample can be obtained by randomly generating a channel matrix $\mathbf{H}$, a transmitted signal $\mathbf{x}$, and a noise vector $\mathbf{z}$. The received signal $\mathbf{y}$ and the channel $\mathbf{H}$ are used to build the weight matrices and the transmitted signal $\mathbf{x}$ is used as the target. After the offline training processing, the trained step sizes $\{\alpha_{\ell}\}$ are ready to be used for the online detection phase. Similar to DetNet for unquantized MIMO detection \cite{Samuel2019Learning}, OBMNet for one-bit MIMO detection does not need to be retrained for a new channel realization $\mathbf{H}$. 

\section{Nearest-Neighbor Search for\\Second-Stage Detection}
\label{sec_nearest_neighbors_search}
Given a received signal, as discussed above we can either use a linear receiver or OBMNet to obtain an estimate $\tilde{\mathbf{x}}$ of the transmitted signal $\mathbf{x}$. However, these receivers all ignore the constraint that the transmitted signal $\mathbf{x}$ belongs to a known discrete set of constellation points. Ignoring this constraint can result in elements of the estimate $\tilde{\mathbf{x}}$ that are well removed from the constellation points, and thus detection errors are likely to occur once symbol-by-symbol detection is applied. This motivates us to propose here an NN search method as a second detection stage in order to fine-tune the solution of stage~1.

The proposed NN search method first finds a limited set of symbol vectors that are nearest to $\tilde{\mathbf{x}}$ and then searches over that set for the most likely symbol vector as the final detection solution. As mentioned in the Introduction, this idea has already been used in~\cite{choi2016near} and~\cite{Nguyen2020SVM}. However, the search space for the methods in~\cite{choi2016near} and~\cite{Nguyen2020SVM} is very large when the number of users is large, and so they are not efficient in terms of computational complexity. The contribution of the proposed NN search method is that it generates searches over a limited number of symbol vectors that are nearest to the estimate $\tilde{\mathbf{x}}$, and thus significantly reduces the computational load.

We denote $\mathcal{M}$ as the constellation in the real domain; for example, $\mathcal{M} = \left\{\pm \frac{1}{\sqrt{2}} \right\}$ for QPSK and $\mathcal{M} = \left\{\pm \frac{1}{\sqrt{10}}, \pm \frac{3}{\sqrt{10}} \right\}$ for 16-QAM. Let $\mathcal{B}$ be the set of decision boundary points; {\em i.e.,} $\mathcal{B} =\{0\}$ for QPSK and $\mathcal{B} =\left\{0,\pm\frac{2}{\sqrt{10}}\right\}$ for 16-QAM. Denote $\tilde{\mathbf{x}} = [\tilde{x}_1,\ldots,\tilde{x}_{2K}]^T$ and $\mathbf{b} = [b_1,\ldots,b_{2K}]^T$, where $b_i$ is the decision boundary point that is nearest to $\tilde{x}_i$, as follows:
\begin{equation}
	b_i = \argmin_{b\in\mathcal{B}}\; |b-\tilde{x}_i|, \quad i \in \{1,2,\ldots,2K\}.
\end{equation}

\begin{figure}[t!]
	\centering
	\begin{tikzpicture}
	\draw [semithick,->] (-3.4,6.5) to (3.4,6.5);
	\draw [semithick] (0,6.45) node[above]{0} to (0,6.55);
	\draw [fill] (2.1213,6.5)  node[below]{$\frac{1}{\sqrt{2}}$} circle [radius=0.05];
	\draw [fill] (-2.1213,6.5) node[below]{$\frac{-1}{\sqrt{2}}$}circle [radius=0.05];
	\draw [fill=red] (1.8,6.5) node[above]{$\tilde{x}_i$}circle [radius=0.05];
	\draw [thin,->] (-0.5,7) to (-0.15,6.8);
	\node [left] at (-0.45,7.1) {$b_i$};
	
	\draw [semithick,->] (-3.4,4) to (3.4,4);
	\draw [semithick] (0,3.95) node[below]{0} to (0,4.05);
	\draw [semithick] (1.8974,3.95) node[below]{$\frac{2}{\sqrt{10}}$} to (1.8974,4.05);
	\draw [semithick] (-1.8974,3.95) node[below]{$\frac{-2}{\sqrt{10}}$} to (-1.8974,4.05);
	
	\draw [fill] (0.9487,4) node[below]{$\frac{1}{\sqrt{10}}$} circle [radius=0.05];
	\draw [fill] (-0.9487,4) node[below]{$\frac{-1}{\sqrt{10}}$} circle [radius=0.05];
	\draw [fill] (2.8460,4) node[below]{$\frac{3}{\sqrt{10}}$} circle [radius=0.05];
	\draw [fill] (-2.8460,4) node[below]{$\frac{-3}{\sqrt{10}}$} circle [radius=0.05];
	\draw [fill=red] (-2.05,4) node[above]{$\tilde{x}_i$}circle [radius=0.05];
	
	\node at (0,5.2) {\small{decision boundary points}};	
	\draw [dashed,thin,->] (0,5.45) to (0,6.4);
	\draw [dashed,thin,->] (0,5) to (0,4.1);
	\draw [dashed,thin,->] (-0.2,5) to (-1.8574,4.1);
	\draw [dashed,thin,->] (0.2,5) to (1.8574,4.1);
	\draw [thin,->] (-1.8974,2.9) to (-1.8974,3.27);
	\node at (-1.8974,2.7) {$b_i$};
	
	\node [left] at (-3.735,6.5) {\small{QPSK}};
	\node [left] at (-4.68,6.5) {\small{(a)}};
	\node [left] at (-3.4,4) {\small{16-QAM}};
	\node [left] at (-4.68,4) {\small{(b)}};
	\end{tikzpicture}
	\caption{An example for the relative difference between $\tilde{x}_i$ and the constellation points: (a) the estimate $\tilde{x}_i$ is far from $b_i=0$ and close to the constellation point $1/\sqrt{2}$, which means there is a high probability that the transmitted signal $x_i$ is $1/\sqrt{2}$; (b) the estimate $\tilde{x}_i$ is close to the boundary point $b_i = -2/\sqrt{10}$, thus it is difficult to say if $-3/\sqrt{10}$ or $-1/\sqrt{10}$ was transmitted.}
	\label{fig_xtilde_vs_constellation_points}
\end{figure}
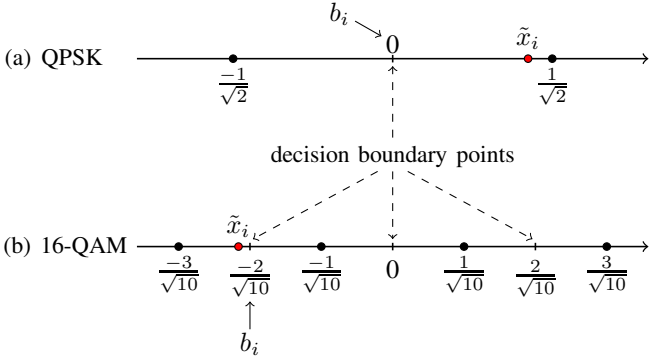
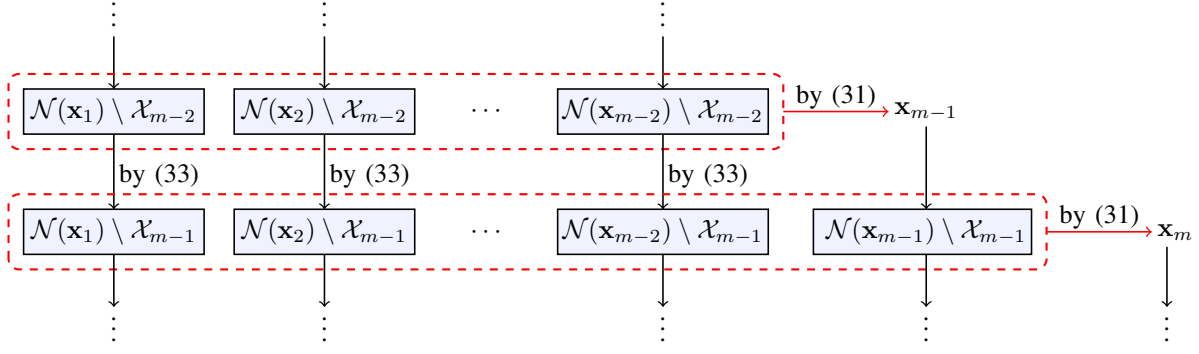
\begin{figure*}[t!]
	\centering
	\begin{tikzpicture}
	
	
	
	\node at (1.7,7.5) {$\vdots$};
	\node at (4.5,7.5) {$\vdots$};
	\node at (9,7.5) {$\vdots$};
	
	\draw [semithick,->] (1.7,7.1) to (1.7,6.4);
	\draw [semithick,->] (4.5,7.1) to (4.5,6.4);
	\draw [semithick,->] (9,7.1) to (9,6.4);
	
	\draw [dashed, thick, red, rounded corners] (0.3,5.6) rectangle (10.6,6.6);
	\draw [fill=InOutColor, semithick] (0.5,5.8) rectangle (2.9,6.4);
	\node at (1.7,6.1) {$\mathcal{N}(\mathbf{x}_1)\setminus \mathcal{X}_{m-2}$};
	\draw [fill=InOutColor, semithick] (3.3,5.8) rectangle (5.7,6.4);
	\node at (4.5,6.1) {$\mathcal{N}(\mathbf{x}_2)\setminus \mathcal{X}_{m-2}$};
	\node at (6.675,6.1) {$\cdots$};
	\draw [fill=InOutColor, semithick] (7.6,5.8) rectangle (10.4,6.4);
	\node at (9,6.1) {$\mathcal{N}(\mathbf{x}_{m-2})\setminus \mathcal{X}_{m-2}$};
	\draw [semithick, red, ->] (10.6,6.1) to (12,6.1);
	\node at (12.5,6.1) {$\mathbf{x}_{m-1}$};
	\node at (11.3,6.3) {by~\eqref{eq_find_xm_using_NCmm1}};
	
	\draw [semithick,->] (1.7,5.8) to (1.7,4.8);
	\node at (2.3,5.25) {by~\eqref{eq_recursive_N}};
	\draw [semithick,->] (4.5,5.8) to (4.5,4.8);
	\node at (5.1,5.25) {by~\eqref{eq_recursive_N}};
	\draw [semithick,->] (9,5.8) to (9,4.8);
	\node at (9.6,5.25) {by~\eqref{eq_recursive_N}};
	\draw [semithick,->] (12.5,5.9) to (12.5,4.8);
	
	\draw [dashed, thick, red, rounded corners] (0.3,4) rectangle (14.1,5);
	\draw [fill=InOutColor, semithick] (0.5,4.2) rectangle (2.9,4.8);
	\node at (1.7,4.5) {$\mathcal{N}(\mathbf{x}_1)\setminus \mathcal{X}_{m-1}$};
	\draw [fill=InOutColor, semithick] (3.3,4.2) rectangle (5.7,4.8);
	\node at (4.5,4.5) {$\mathcal{N}(\mathbf{x}_2)\setminus \mathcal{X}_{m-1}$};
	\node at (6.675,4.5) {$\cdots$};
	\draw [fill=InOutColor, semithick] (7.6,4.2) rectangle (10.4,4.8);
	\node at (9,4.5) {$\mathcal{N}(\mathbf{x}_{m-2})\setminus \mathcal{X}_{m-1}$};
	\draw [fill=InOutColor, semithick] (11,4.2) rectangle (13.9,4.8);
	\node at (12.5,4.5) {$\mathcal{N}(\mathbf{x}_{m-1})\setminus \mathcal{X}_{m-1}$};
	\draw [semithick, red, ->] (14.1,4.5) to (15.5,4.5);
	\node at (14.8,4.7) {by~\eqref{eq_find_xm_using_NCmm1}};
	\node [right]  at (15.45,4.5) {$\mathbf{x}_m$};
	
	\draw [semithick,->] (1.7,4.2) to (1.7,3.5);
	\draw [semithick,->] (4.5,4.2) to (4.5,3.5);
	\draw [semithick,->] (9,4.2) to (9,3.5);
	\draw [semithick,->] (12.5,4.2) to (12.5,3.5);
	\draw [semithick,->] (15.7,4.3) to (15.7,3.5);
	
	\node [below] at (1.7,3.7) {$\vdots$};
	\node [below] at (4.5,3.7) {$\vdots$};
	\node [below] at (9,3.7) {$\vdots$};
	\node [below] at (12.5,3.7) {$\vdots$};
	\node [below] at (15.7,3.7) {$\vdots$};
	\end{tikzpicture}
	\caption{Flowchart of the proposed nearest-neighbor search method. A recursive formation of sets is exploited to reduce the computational complexity. A subset $\mathcal{N}(\mathbf{x}_{p})\setminus \mathcal{X}_{m-1}$ with $p\in \{1,\ldots,m-2\}$ is obtained by removing $\mathbf{x}_{m-1}$ from the subset $\mathcal{N}(\mathbf{x}_{p})\setminus \mathcal{X}_{m-2}$ as given in~\eqref{eq_recursive_N}. The last subset $\mathcal{N}(\mathbf{x}_{m-1})\setminus \mathcal{X}_{m-1}$ is obtained by using $\mathbf{x}_{m-1}$ and other nearest symbol vectors. The $m^\text{th}$ nearest symbol vector $\mathbf{x}_m$ is then obtained by searching over the $m-1$ subsets.}
	\label{fig_flow_chart}
\end{figure*}

An illustrative example for the relative difference between $\tilde{x}_i$ and the constellation points is given in Fig.~\ref{fig_xtilde_vs_constellation_points}. This example illustrates the problem that occurs when $\tilde{x}_i$ is close to a decision boundary point, where symbol-by-symbol detection may not be reliable. Here, we use a threshold $\gamma>0$ to classify whether symbol-by-symbol detection is used or not. More specifically, if the distance from $\tilde{x}_i$ to its nearest decision boundary point $b_i$ is greater than $\gamma$, i.e., $|\tilde{x}_i-b_i|>\gamma$, then we can use symbol-by-symbol detection for $\tilde{x}_i$. When $|\tilde{x}_i-b_i|\leq\gamma$, symbol-by-symbol detection is not reliable, and so we list the two nearest constellation points to $\tilde{x}_i$ as the candidates for the transmitted signal $x_i$.

Let $\mathcal{A}_i$ denote the set of candidates for the transmitted signal $x_i$. When $|\tilde{x}_i-b_i|>\gamma$, we apply symbol-by-symbol detection and so 
\begin{equation*}
\mathcal{A}_i = \left\{\argmin_{x\in\mathcal{M}}\, |x-\tilde{x}_i|\right\}.
\end{equation*}
When $|\tilde{x}_i-b_i|\leq\gamma$, we have $\mathcal{A}_i = \left\{b_i \pm \frac{1}{\sqrt{2}} \right\} = \left\{\pm\frac{1}{\sqrt{2}} \right\}$ for QPSK and $\mathcal{A}_i = \left\{b_i \pm \frac{1}{\sqrt{10}} \right\}$ for 16-QAM. Hence, $\mathcal{A}_i$ contains only one or two elements. The following example illustrates the formation of $\mathcal{A}_i$.

\begin{example}
	Suppose that $\tilde{\mathbf{x}} = [0.1, -0.5, -0.3, 0.8]^T$ and QPSK modulation is used with $\gamma = \frac{1}{2\sqrt{2}} \approx 0.35$. Note here that $b_1 = b_2 = b_3 = b_4 = 0$. We have
	\begin{itemize}
		\item $\mathcal{A}_1 = \mathcal{A}_3 = \big\{\pm \frac{1}{\sqrt{2}}\big\}$ because $|\tilde{x}_1 - b_1| = 0.1 < \gamma$ and $|\tilde{x}_3 - b_3| = 0.3 < \gamma$,
		\item $\mathcal{A}_2 = \big\{\frac{-1}{\sqrt{2}}\big\}$ because $|\tilde{x}_2 - b_2| = 0.5 > \gamma$ and $\tilde{x}_2$ is closer to $\frac{-1}{\sqrt{2}}$ than $\frac{1}{\sqrt{2}}$, i.e., $\big|\tilde{x}_2 - \frac{-1}{\sqrt{2}}\big| < \big|\tilde{x}_2 - \frac{1}{\sqrt{2}}\big|,$
		\item $\mathcal{A}_4 = \big\{\frac{1}{\sqrt{2}}\big\}$ because $|\tilde{x}_4 - b_4| = 0.8 > \gamma$ and $\tilde{x}_4$ is closer to $\frac{1}{\sqrt{2}}$ than $\frac{-1}{\sqrt{2}}$, i.e., $\big|\tilde{x}_4 - \frac{1}{\sqrt{2}}\big| < \big|\tilde{x}_4- \frac{-1}{\sqrt{2}}\big|.$
	\end{itemize}
	Hence, in this example, $\mathcal{A}_1$ and $\mathcal{A}_3$ have two elements while $\mathcal{A}_2$ and $\mathcal{A}_4$ have only one element.
\end{example}

The complete set of candidates for the transmitted signal vector is given by the Cartesian product $$\mathcal{A} = \mathcal{A}_1\times \mathcal{A}_2\times\ldots\times\mathcal{A}_{2K} ,$$ and so the size of $\mathcal{A}$ is $|\mathcal{A}| = \prod_{i=1}^{2K}|\mathcal{A}_i| = 2^A$, where $A$ is the number of sets $\mathcal{A}_i$ having two elements. The existing search methods in~\cite{choi2016near} and~\cite{Nguyen2020SVM} always search over the entire set $\mathcal{A}$. However, it can be seen that the size of $\mathcal{A}$ grows exponentially with $A$. In addition, $A$ also grows as the number of users $K$ increases. Thus, searching over the entire list $\mathcal{A}$ as in~\cite{choi2016near} and~\cite{Nguyen2020SVM} can be prohibitively complex when the number of users is large. 

On the other hand, the proposed NN search method finds a set of $M$ symbol vectors in $\mathcal{A}$ that are nearest to $\tilde{\mathbf{x}}$, then searches over that smaller set for the final solution. In this way, the NN search method can limit the computational complexity. Note that a symbol vector in this context is any element of $\mathcal{A}$. Let $\mathcal{X}_M = \{\mathbf{x}_1,\mathbf{x}_2,\ldots,\mathbf{x}_M\}$ denote the set of the $M$ nearest symbol vectors to $\tilde{\mathbf{x}}$. The larger $M$ is, the higher the probability that the set $\mathcal{X}_M$ contains the true symbol vector. However, a large value of $M$ will result in more computation for the search. Therefore, $M$ should be chosen to achieve a good trade-off between detection accuracy and computational complexity. The value of $M$ can be chosen by empirical evaluations. The main challenge here is how to find the $M$ nearest symbol vectors to $\tilde{\mathbf{x}}$ quickly and efficiently. To address this problem, we employ the following notation and definitions.

For any two symbol vectors $\mathbf{x}\in \mathcal{A}$ and $\mathbf{x}'\in\mathcal{A}$, let $d(\mathbf{x}, \mathbf{x}')$ denote the number of position indices at which the elements of $\mathbf{x}$ are different from the corresponding elements of $\mathbf{x}'$. Since each element of $\mathbf{x}$ and $\mathbf{x}'$ belongs to a finite set of just one or two elements, $d(\mathbf{x}, \mathbf{x}')$ is actually the Hamming distance between $\mathbf{x}$ and $\mathbf{x}'$.

\begin{definition}[Neighbor of a symbol vector]
	A symbol vector $\mathbf{x}$ is called a neighbor of another symbol vector $\mathbf{x}'$, or vice versa, when the Hamming distance between them is one, i.e., $d(\mathbf{x}, \mathbf{x}')=1$.
\end{definition}

\begin{definition}[Neighbor of a set]
	Given a set of symbol vectors $\mathcal{S}$ and another symbol vector $\mathbf{x}\notin\mathcal{S}$, let
	\begin{equation}
		d_\mathrm{min}(\mathbf{x},\mathcal{S}) = \min_{\mathbf{x}'\in\mathcal{S}} d(\mathbf{x},\mathbf{x}').
	\end{equation}
	The symbol vector $\mathbf{x}$ is called a neighbor of $\mathcal{S}$ if and only if $d_\mathrm{min}(\mathbf{x},\mathcal{S}) = 1$, or in other words, if and only if $\mathbf{x}$ is the neighbor of at least one member of $\mathcal{S}$.
\end{definition}

Let $\mathcal{N}(\mathbf{x})$ and $\mathcal{N}(\mathcal{S})$ denote the set of neighbors of symbol vector $\mathbf{x}$ and set $\mathcal{S}$, respectively. Let $\mathcal{X}_M = \left\{\mathbf{x}_1,\mathbf{x}_2,\ldots,\mathbf{x}_M\right\}$ with $\mathbf{x}_m \in \mathcal{A}$ and $m\in\{1,2,\ldots,M\}$ denote the set of the $M$ nearest symbol vectors to $\tilde{\mathbf{x}}$ satisfying
\begin{equation}
	\|\mathbf{x}_1-\tilde{\mathbf{x}}\|^2 < \|\mathbf{x}_2-\tilde{\mathbf{x}}\|^2 <\ldots< \|\mathbf{x}_M-\tilde{\mathbf{x}}\|^2 < \|\mathbf{x}-\tilde{\mathbf{x}}\|^2
\end{equation}
where $\mathbf{x}$ is any symbol vector in $\mathcal{A}$, but not in $\mathcal{X}_M$. Hence, $\mathbf{x}_m$ is the $m^\mathrm{th}$ nearest symbol vector to $\tilde{\mathbf{x}}$. Clearly, the nearest symbol vector $\mathbf{x}_1$ is obtained by applying symbol-by-symbol detection to $\tilde{\mathbf{x}}$. The problem now is how to efficiently find $\mathbf{x}_2$, \ldots, $\mathbf{x}_M$. The following proposition can be exploited to solve this problem.
\begin{proposition}
	\label{proposition1}
	The $m^\mathrm{th}$ nearest symbol vector $\mathbf{x}_m$ must be a neighbor of the set $\mathcal{X}_{m-1} = \{\mathbf{x}_1,\mathbf{x}_2,\ldots,\mathbf{x}_{m-1}\}$, i.e.,
	\begin{equation*}
		\mathbf{x}_m \in \mathcal{N}(\mathcal{X}_{m-1}).
	\end{equation*}
\end{proposition}
\begin{IEEEproof}
	Please refer to Appendix~\ref{append1}
\end{IEEEproof}

Proposition~\ref{proposition1} indicates that we can find the $m^{\mathrm{th}}$ nearest symbol vector $\mathbf{x}_m$ from the neighbor set of $\mathcal{X}_{m-1}$, i.e.,
\begin{equation}
\mathbf{x}_{m} = \argmin_{\mathbf{x}\in \mathcal{N}\left(\mathcal{X}_{m-1}\right)} \|\mathbf{x}-\tilde{\mathbf{x}}\|^2
\label{eq_find_xm_using_NCmm1}
\end{equation}
where $\mathcal{N}\left(\mathcal{X}_{m-1}\right)$ is the neighbor set of $\mathcal{X}_{m-1}$ and is given as
\begin{align}
	\mathcal{N}\left(\mathcal{X}_{m-1}\right) & = \bigg(\bigcup_{p=1}^{m-1} \mathcal{N}\left(\mathbf{x}_p\right)\bigg)\setminus \mathcal{X}_{m-1} \notag\\
	& = \bigcup_{p=1}^{m-1} \Big(\mathcal{N}\left(\mathbf{x}_p\right)\setminus \mathcal{X}_{m-1}\Big).
	\label{eq_neighbor_set_of_Cmm1}
\end{align}

Hence, in order to find $\mathbf{x}_m$, we need to accomplish two tasks: (i) find $m-1$ subsets $\{\mathcal{N}(\mathbf{x}_p)\setminus \mathcal{X}_{m-1}\}_{p=1,\ldots,m-1}$ and (ii) search for $\mathbf{x}_m$ within the subsets. The method of directly finding the $m-1$ subsets and then searching them for $\mathbf{x}_m$ is not efficient. In the following, we present a recursive strategy to obtain $\mathbf{x}_m$ quickly and efficiently.

Note that the inner term on the right-hand side of~\eqref{eq_neighbor_set_of_Cmm1} can be written as follows:
\begin{equation}
	\mathcal{N}(\mathbf{x}_p)\setminus \mathcal{X}_{m-1} =  \Big(\mathcal{N}(\mathbf{x}_p)\setminus \mathcal{X}_{m-2}\Big) \setminus \{\mathbf{x}_{m-1}\}.
	\label{eq_recursive_N}
\end{equation}
Therefore, we can exploit~\eqref{eq_recursive_N} to obtain the first $m-2$ subsets $\{\mathcal{N}(\mathbf{x}_p)\setminus \mathcal{X}_{m-1}\}_{p=1,\ldots,m-2}$ by removing $\mathbf{x}_{m-1}$ from $m-2$ other subsets $\{\mathcal{N}(\mathbf{x}_p)\setminus \mathcal{X}_{m-2}\}_{p=1,\ldots,m-2}$, which were already obtained previously when we found $\mathbf{x}_{m-1}$. The last subset $\mathcal{N}(\mathbf{x}_{m-1})\setminus\mathcal{X}_{m-1}$ is obtained by using $\mathbf{x}_{m-1}$ and the other nearest symbol vectors. A flowchart illustrating this recursive strategy is given in Fig.~\ref{fig_flow_chart}.
\begin{algorithm}[t!]
	\small
	\KwIn{$\tilde{\mathbf{x}}$, $\gamma$, $M$.}
	\KwOut{$\hat{\mathbf{x}}$.}
	Find $\mathbf{b}$ and $\mathcal{A}_1, \mathcal{A}_2, \ldots, \mathcal{A}_{2K}$ based on $\mathbf{b}$\;
	Let $|\mathcal{A}| = \prod_{i=1}^{2K}|\mathcal{A}_i|$\;
	\eIf{$|\mathcal{A}|\leq M$}{
		Let $\mathcal{A} = \mathcal{A}_1\times\mathcal{A}_2\times\ldots\times\mathcal{A}_{2K}$\;
		$\hat{\mathbf{x}} = \argmin_{\mathbf{x}\in \mathcal{A}} \mathcal{P}_\mathrm{robust}(\mathbf{x})$\;
	}{
		Find $\mathbf{x}_1$ via symbol-by-symbol detection\;
		Let $\mathcal{C}_{1} = \operatorname{sort}\left(\mathcal{N}(\mathbf{x}_1)\right)$\label{algo1_sorting_x1}\;
		\For{$m=2$ \textup{\textbf{to}} $M$}{
			Let $\mathcal{S}_m = \{\mathcal{C}_{1}[1], \mathcal{C}_{2}[1], \ldots, \mathcal{C}_{m-1}[1]\}$\label{algo1_Sm}\;
			$\mathbf{x}_m = \argmin_{\mathbf{x} \in \mathcal{S}_m} \|\mathbf{x} - \tilde{\mathbf{x}}\|^2$\label{algo1_xm}\;
			\If{$m<M$}{
				\For{$p=1$ \textup{\textbf{to}} $m-1$}{\label{algo1_start_remove_xm}
					\If {$\mathcal{C}_{p}[1] = \mathbf{x}_m$} {
						Remove $\mathcal{C}_{p}[1]$ from $\mathcal{C}_{p}$\;
					}
				}\label{algo1_stop_remove_xm}
				
				Let $\mathcal{C}_{m} = \operatorname{sort}\left(\mathcal{N}\left(\mathbf{x}_{m}\right)\right)$\label{algo1_sorting_xm}\;
				\For{$p=1$ \textup{\textbf{to}} $m-1$}{
					\If {$\mathcal{C}_{m}[1] = \mathbf{x}_p$} {
						Remove $\mathcal{C}_{p}[1]$ from $\mathcal{C}_{m}$\;
					}
				}\label{algo1_stop_forming_last_subset}
			}
		}
		$\hat{\mathbf{x}} = \argmin_{\mathbf{x}\in \mathcal{X}_M} \mathcal{P}_\mathrm{robust}(\mathbf{x})$\label{algo1_final_solution}\;
	}
	\Return{$\hat{\mathbf{x}}$}\;
	\caption{Proposed Nearest-Neighbors Search.}
	\label{algo_nearest_neighbors_search}
\end{algorithm}

\textit{Remark 1:} If the elements of $\mathcal{N}(\mathbf{x}_p)\setminus \mathcal{X}_{m-2}$ are already sorted in ascending order of distance to $\tilde{\mathbf{x}}$, then $\mathbf{x}_{m-1}$ can be removed from $\mathcal{N}(\mathbf{x}_p)\setminus \mathcal{X}_{m-2}$ by simply checking the first element of $\mathcal{N}(\mathbf{x}_p)\setminus \mathcal{X}_{m-2}$. The reason for this is that $\mathbf{x}_{m-1}$ is the $(m-1)^\text{th}$ nearest symbol vector, which means the distance from $\mathbf{x}_{m-1}$ to $\tilde{\mathbf{x}}$ cannot be greater than the distance from any element of $\mathcal{N}(\mathbf{x}_p)\setminus \mathcal{X}_{m-2}$ to $\tilde{\mathbf{x}}$. In addition, the elements of $\mathcal{N}(\mathbf{x}_p)\setminus \mathcal{X}_{m-2}$ are distinct and already sorted, and so if $\mathbf{x}_{m-1}$ exists in $\mathcal{N}(\mathbf{x}_p)\setminus \mathcal{X}_{m-2}$, it must be the first element of $\mathcal{N}(\mathbf{x}_p)\setminus \mathcal{X}_{m-2}$.

\textit{Remark 2:} If the elements of each subset $\mathcal{N}(\mathbf{x}_p)\setminus \mathcal{X}_{m-1}$ are already sorted in ascending order of distance to $\tilde{\mathbf{x}}$, then the search over the $m-1$ subsets for $\mathbf{x}_m$ can be done by simply searching over a list of $m-1$ candidates, where each candidate is the first element of a subset $\mathcal{N}(\mathbf{x}_p)\setminus \mathcal{X}_{m-1}$.

Based on the observations in Remarks~1 and~2, we propose the nearest-neighbor search method described in Algorithm~\ref{algo_nearest_neighbors_search}. The key idea is to use the recursive strategy depicted in Fig.~\ref{fig_flow_chart} and to implement the observations made in Remarks~1 and~2. Whenever forming a set $\mathcal{N}(\mathbf{x}_m)$, we sort its elements in ascending order of distance to $\tilde{\mathbf{x}}$ as described in lines~\ref{algo1_sorting_x1} and~\ref{algo1_sorting_xm} of Algorithm~\ref{algo_nearest_neighbors_search}. In this way, we only need to sort $M-1$ times, and the remainder of the proposed algorithm only involves comparisons based on checking the first elements of the subsets. We denote $\mathcal{C}_1, \ldots, \mathcal{C}_{M-1}$ as the subsets corresponding to $\mathbf{x}_1, \ldots, \mathbf{x}_{M-1}$, respectively, and $\mathcal{C}_m[1]$ denotes the first element of the subset $\mathcal{C}_m$. Lines~\ref{algo1_Sm} and~\ref{algo1_xm} implement Remark~2 to obtain $\mathbf{x}_m$. Remark~1 is implemented in lines~\ref{algo1_start_remove_xm}-\ref{algo1_stop_remove_xm}. The last subset is obtained in lines~\ref{algo1_sorting_xm}-\ref{algo1_stop_forming_last_subset}. Finally, line~\ref{algo1_final_solution} gives the final solution by searching for the highest-likelihood symbol vector among the $M$ nearest symbol vectors. 

\section{Computational Complexity Analysis and Numerical Results}
\label{sec_complexity_analysis_and_numerical_results}

\subsection{Computational Complexity Analysis}
\label{subsec_complexity_analysis}
\begin{table}[t!]
	\centering
	\caption{Computational Complexity Comparison: $T_\mathrm{d}$ is the data block length, $\kappa(N)$ is a super-linear function of $N$, and $GN_\mathrm{s}=2N$.\label{table_0}}
	\renewcommand{\arraystretch}{1.4}
	\begin{tabular}{|l|c|l|}
		\hline
		\textbf{Method}& \textbf{Preprocessing} & \textbf{Stage 1} \\
		\hline
		\textbf{BMRC} & $\mathcal{O}(KN)$ & \multirow{3}{*}{$\mathcal{O}(KNT_\mathrm{d})$} \\ \cline{1-2}
		\textbf{BZF} & $\mathcal{O}(K^2N)$ &  \\ \cline{1-2}
		\textbf{BMMSE} & $\mathcal{O}(\max\{KN^2,N^{2.373}\})$ & \\ \hline
		\textbf{DNN-based} & -- & $\mathcal{O}(KNLT_\mathrm{d})$ \\ \hline
		\textbf{SVM-based~\cite{Nguyen2020SVM}} & -- & $\mathcal{O}(KN\kappa(N)T_\mathrm{d})$ \\ \hline
		\textbf{OSD}~\cite{Jeon2018One} & $\mathcal{O}(4^{N/G}KN|\bar{\mathcal{M}}|^K)$ & $\mathcal{O}\big((N/N_\mathrm{s})KNT_\mathrm{d}\big)$ \\ \hline
	\end{tabular}
\end{table}
A computational complexity comparison in terms of big-$\mathcal{O}$ notation is provided in Table~\ref{table_0}. It can be seen that the computational complexity of the proposed receivers is lower than that of existing methods. In particular, the linear receivers have the lowest complexity, while the OSD method in~\cite{Jeon2018One} has the highest complexity, which grows exponentially with $K$ and $N$. Note that the complexity of the SVM-based method~\cite{Nguyen2020SVM} is due to the decomposition techniques used to solve the SVM problem, e.g.,~\cite{platt1998sequential,joachims1998making,hsu2002simple}. The term $\kappa(N)$ is empirically reported to be a super-linear function of $N$. The complexity of the DNN-based OBMNet detector is only $\mathcal{O}(KNLT_\mathrm{d})$, which is lower than that of the SVM-based method.
	
The computational complexity of the proposed NN search method is $\mathcal{O}(MK\max\{M,N\}T_\mathrm{d})$ in the worst case. This complexity is mainly due to the detection step for $\hat{\mathbf{x}}$ and the \textbf{for} loops as described in Algorithm~\ref{algo_nearest_neighbors_search}. The complexity of the full $\mathcal{A}$-space search method is $\mathcal{O}(|\mathcal{A}|KNT_\mathrm{d})$ where $|\mathcal{A}|$ can grow exponentially with $K$.

\subsection{Numerical Results}
\label{subsec_numerical_results}
\begin{figure}[t!]
	\centering
	\begin{subfigure}[t]{0.5\textwidth}
		\centering
		\includegraphics[width=\linewidth]{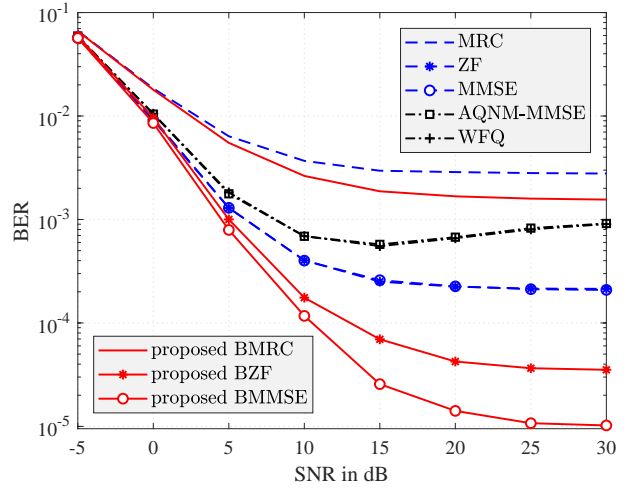}
		\caption{$K = 2$, $N = 16$.}
		\label{fig_6a}
	\end{subfigure}%
	
	\begin{subfigure}[t]{0.5\textwidth}
		\centering
		\includegraphics[width=\linewidth]{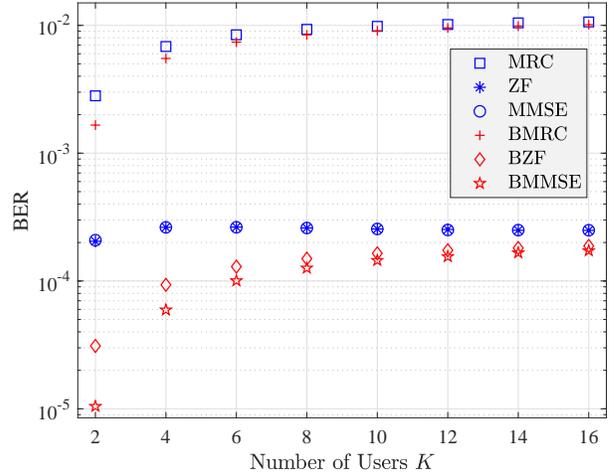}
		\caption{$N = 8K$ and $\rho = 30$ dB.}
		\label{fig_6b}
	\end{subfigure}
	\caption{First stage performance comparison between the proposed and existing linear receivers with QPSK signaling.}
	\label{fig_6}
\end{figure}

\begin{figure*}[t!]
	\centering
	\begin{subfigure}[t]{0.49\textwidth}
		\centering
		\includegraphics[width=\linewidth]{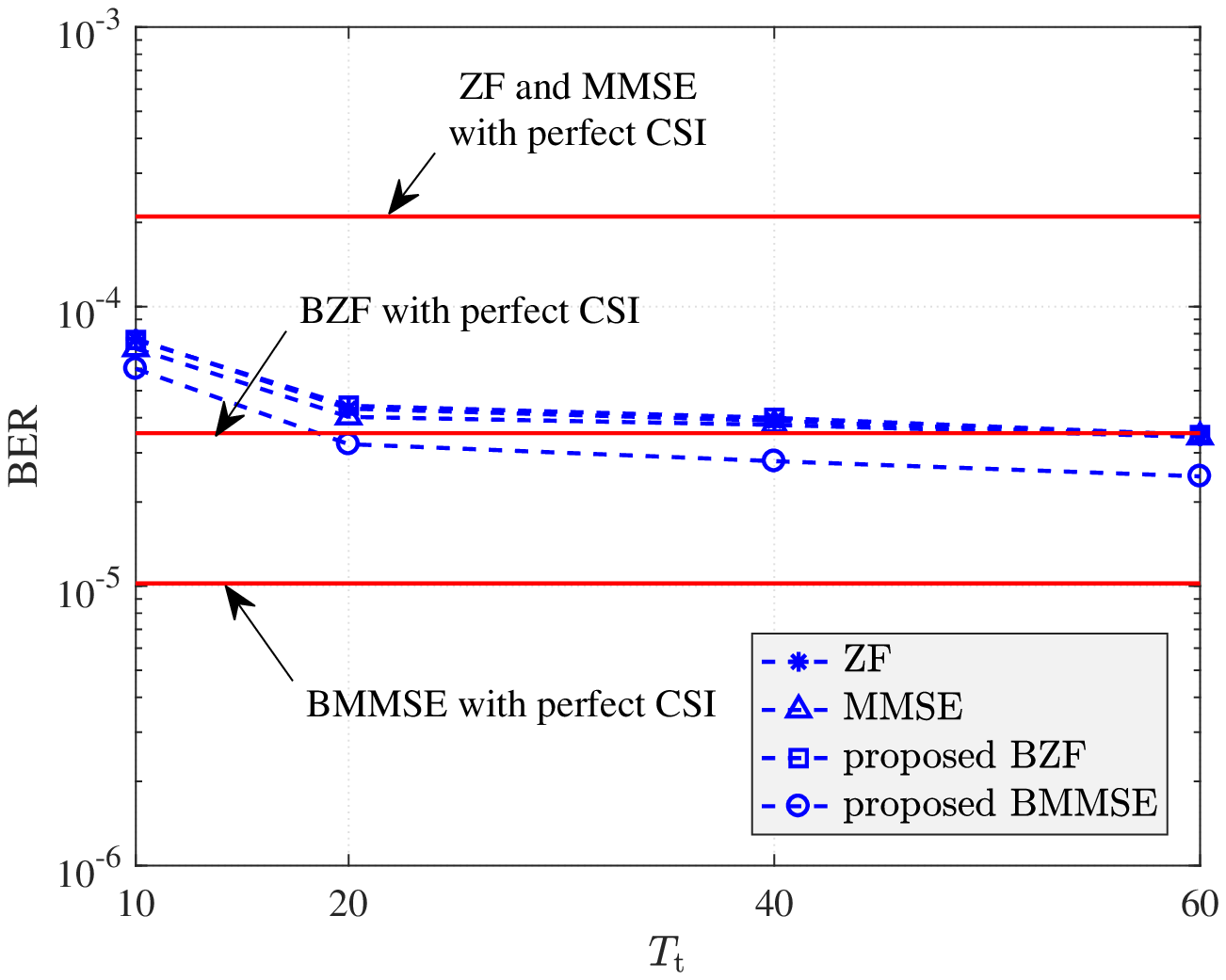}
		\caption{BMMSE estimated CSI~\cite{li2017channel}.}
		\label{fig_linear_receivers_est_CSI_a}
	\end{subfigure}
	~
	\begin{subfigure}[t]{0.49\textwidth}
		\centering
		\includegraphics[width=\linewidth]{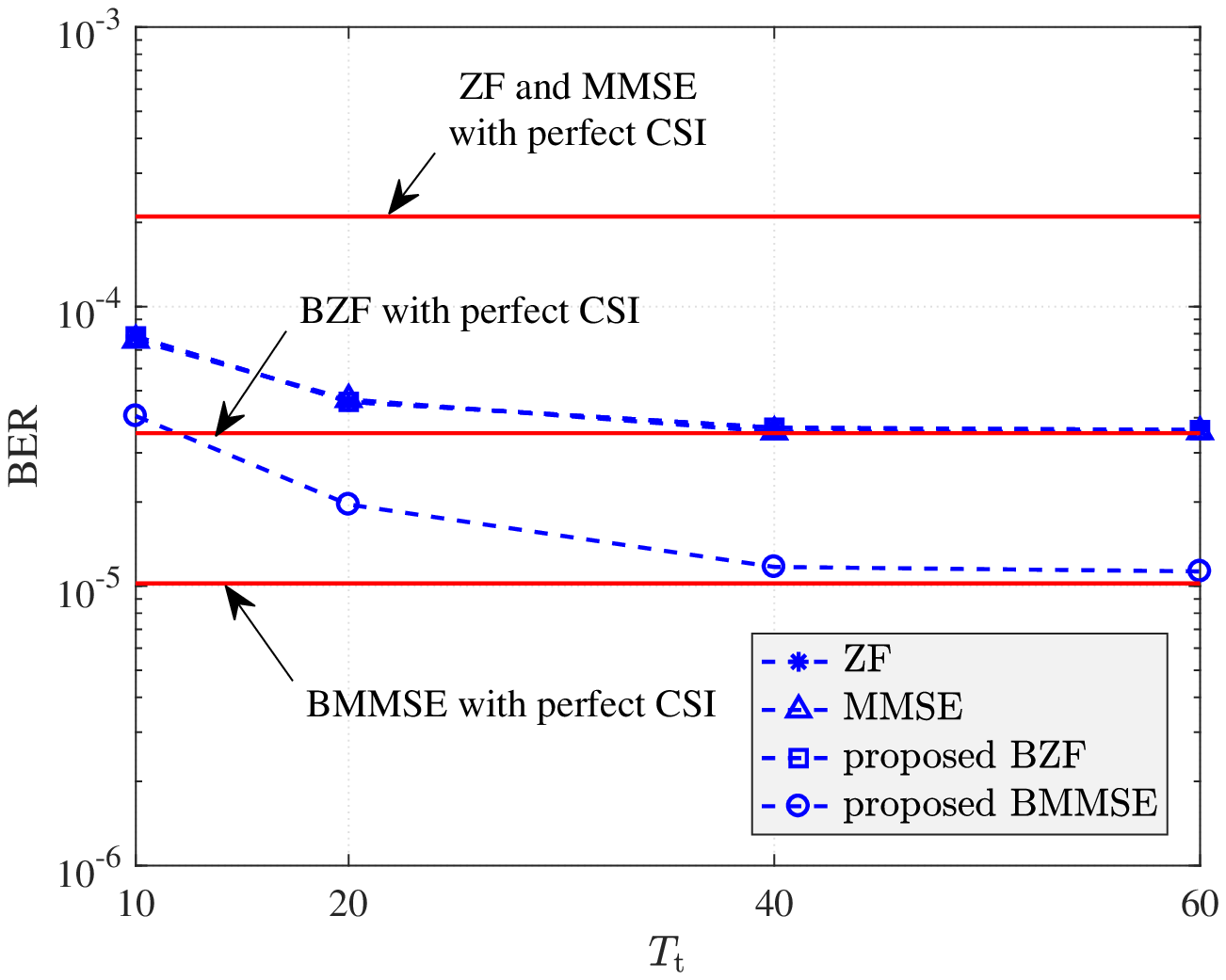}
		\caption{SVM-based estimated CSI~\cite{Nguyen2020SVM}.}
		\label{fig_linear_receivers_est_CSI_b}
	\end{subfigure}
	\caption{BER comparison between ZF, MMSE, BZF, and BMMSE linear receivers with estimated CSI. The setting is $K=2$ users, $N=16$ receive antennas, QPSK signaling, and $\mathrm{SNR}=30\mathrm{dB}$. $T_\mathrm{t}$ is the training length.}
	\label{fig_linear_receivers_est_CSI}
\end{figure*}
This section presents numerical results to show the performance of the proposed two-stage detection methods. The channel elements are assumed to be i.i.d. and each channel element is generated from the normal distribution $\mathcal{CN}(0,1)$. 

First, we evaluate the performance of the conventional and proposed Bussgang-based linear receivers assuming perfect CSI is available (examples with estimated CSI will be given next). Fig.~\ref{fig_6} presents BER comparisons between the proposed and existing linear receivers for QPSK signaling. Among the existing receivers, we see that the ZF and MMSE receivers obtain the same performance (blue curves with symbols), as do 
the AQNM-MMSE \cite{Liu2018Asymptotic} and WFQ receivers \cite{mezghani2007modified} (black curves with symbols). The proposed Bussgang-based linear receivers significantly outperform their conventional counterparts. The high-SNR error floors of the proposed linear receivers are much lower than those of the conventional approaches. These performance improvements are achieved thanks to the exact linear input-output relationship of massive MIMO systems with one-bit ADCs obtained by the Bussgang decomposition. In Fig.~\ref{fig_6b}, we evaluate the performance as the number of users $K$ increases. Here, we omit AQNM-MMSE and WFQ since they are outperformed by ZF and MMSE. It is observed that the proposed linear receivers always yield lower BERs than the standard methods, and the performance improvement is best seen when the number of users $K$ is not too
large. As $K$ increases, the gap between the error floors tend to diminish. This is due to the fact that for large $K$, we have $\mathbf{H}\mathbf{H}^H \approx K\mathbf{I}_N$, which yields $\boldsymbol{\Sigma}_r \approx (K+N_0)\mathbf{I}_N$, $\mathbf{A} \approx \sqrt{\mu}\mathbf{H}$ and $\boldsymbol{\Sigma}_n \approx \big(1-\mu K\big)\mathbf{I}_N$, where $\mu = 2/(\pi (K+N_0))$. These approximations result in Bussgang-based linear receivers that are equivalent to the conventional approaches to within a scaling factor:
\begin{align*}
\mathbf{W}_{\mathtt{BMRC}} &\approx \sqrt{\mu}\mathbf{H}^H,\\
\mathbf{W}_{\mathtt{BZF}} & \approx \sqrt{\frac{1}{\mu}}\Big(\mathbf{H}^H\mathbf{H}\Big)^{-1}\mathbf{H}^H,\\
\mathbf{W}_{\mathtt{BMMSE}} & \approx \sqrt{\frac{1}{\mu}}\mathbf{H}^H\bigg(\mathbf{H}\mathbf{H}^H + \frac{1-\mu K}{\mu}\mathbf{I}_N\bigg)^{-1}.
\end{align*}

In Fig.~\ref{fig_linear_receivers_est_CSI}, we provide BER comparisons between the ZF, MMSE, BZF, and BMMSE linear receivers with estimated CSI for a case with $K=2$ users and $N=16$ antennas. Figure~\ref{fig_linear_receivers_est_CSI}(a) shows results for the Bussgang-based channel estimator in \cite{li2017channel}, while Fig.~~\ref{fig_linear_receivers_est_CSI}(b) employs the SVM-based channel estimator of \cite{Nguyen2020SVM}. It can be seen that the BMMSE receiver always outperforms the others. A striking observation is that ZF and MMSE with estimated CSI outperform ZF and MMSE with perfect CSI. There is a reason for this. Recall that Bussgang-based linear receivers BZF and BMMSE use the effective channel
\begin{equation}
    \bar{\mathbf{A}} = \sqrt{\frac{2}{\pi}}\operatorname{diag}(\bar{\mathbf{H}}\bar{\mathbf{H}}^H+N_0\mathbf{I}_N)^{-1/2}\bar{\mathbf{H}}.
    \label{eq_1}
\end{equation}
Let $\bar{\mathbf{A}}_{i,:}$ and $\bar{\mathbf{H}}_{i,:}$ denote the $i^\mathrm{th}$ row of $\bar{\mathbf{A}}$ and $\bar{\mathbf{H}}$, respectively, then we have
\begin{equation}
    \bar{\mathbf{A}}_{i,:} = \sqrt{\frac{2}{\pi}}\frac{\bar{\mathbf{H}}_{i,:}}{\sqrt{\|\bar{\mathbf{H}}_{i,:}\|^2+N_0}}, \quad i = 1,2,\ldots,N.
    \label{eq_2}
\end{equation}
This indicates that the effective channel $\bar{\mathbf{A}}_{i,:}$ is a normalized version of the true channel. Note that the instantaneous magnitude of $\bar{\mathbf{H}}_{i,:}$ is not identifiable in $1$-bit quantized MIMO systems~\cite{Rao2019Channel}, and consequently the SVM-based~\cite{Nguyen2020SVM} and BMMSE~\cite{li2017channel} channel estimators provide estimates whose magnitudes are normalized. Therefore, when using a channel estimator such as \cite{li2017channel,Nguyen2020SVM}, ZF with estimated CSI will give the same performance as BZF with estimated CSI. ZF with estimated CSI outperforms ZF with perfect CSI since the channel estimate takes into account the inherent scaling ambiguity in the observed data. For the same reason, MMSE and BMMSE with estimated CSI also outperform MMSE with perfect CSI, but MMSE performs worse than BMMSE because MMSE still applies the noise covariance matrix $N_0\mathbf{I}$, while BMMSE uses the covariance matrix ${\boldsymbol{\Sigma}}_{\bar{n}}$ that includes information about the quantization noise.

\begin{figure}[t!]
	\centering
	\includegraphics[width=\linewidth]{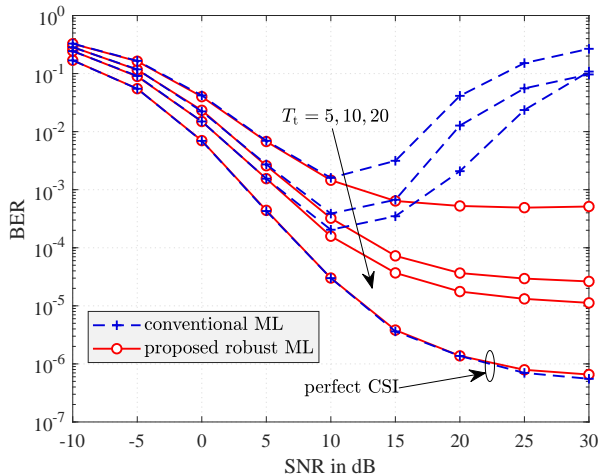}
	\caption{Performance comparison between the conventional and the proposed ML detection problems with $K=2$, $N=16$, and QPSK signaling. The BMMSE channel estimator~\cite{li2017channel} is used with different training lengths $T_\mathrm{t}$.}
	\label{fig_7}
\end{figure}

For the first stage, besides the Bussgang-based linear receiver, we also proposed OBMNet, which is devised from a reformulated robust ML detection problem. In Fig.~\ref{fig_7}, we verify the robustness of the reformulated ML detection problem in~\eqref{eq_robust_ML_detection} when implemented with estimated CSI. We carried out simulations using the BMMSE channel estimator~\cite{li2017channel} with different training lengths $T_\mathrm{t}$. It can be seen from Fig.~\ref{fig_7} that when the CSI is perfectly known, both the conventional and the proposed ML detection algorithms yield almost identical performance. However, when the CSI is imperfectly known, the performance of conventional ML detection is significantly degraded at high SNR, while the proposed robust ML detection algorithm remains stable. This verifies our analysis in Section~\ref{sec_robust_ML_and_DNN_receivers}.
\begin{figure}[t!]
	\centering
	\includegraphics[width=\linewidth]{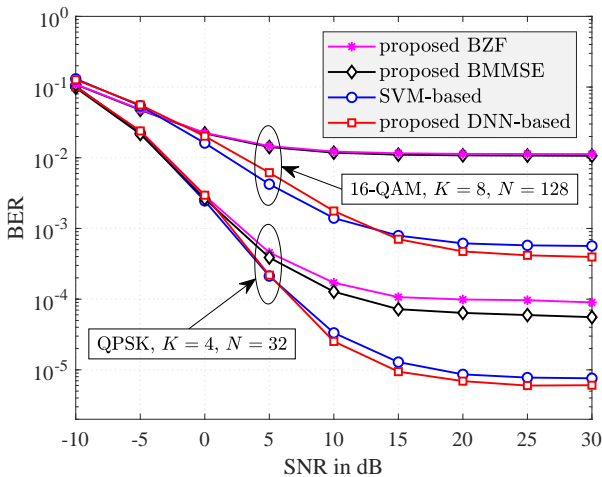}
	\caption{First stage performance comparison between the proposed BZF, BMMSE, and DNN-based receivers and the SVM-based method~\cite{Nguyen2020SVM}.}
	\label{fig_8}
\end{figure}

Fig.~\ref{fig_8} provides a performance comparison between the proposed DNN-based, BMMSE, and BZF receivers and the SVM-based receiver in~\cite{Nguyen2020SVM}. The performance of OSD is comparable to that of the SVM-based method but with much higher computational complexity. Since the SVM-based method also outperforms other prior methods, we use it as a comparative benchmark in this paper. To implement the SVM-based receiver, we use the Scikit-learn machine learning library~\cite{scikit-learn}, and the maximum number of iterations is set to be $30$. For training OBMNet, we use TensorFlow~\cite{tensorflow} and the Adam optimizer~\cite{kingma2014adam} with a learning rate of $10^{-2}$. The size of each training batch is set to $1000$. The input of the first layer $\mathbf{x}_0$ is set to a zero vector. During the detection phase, the trained OBMNet is employed to perform batch detection. Note that batch detection is an advantage of DNN since it can take a batch of multiple symbol vectors as its input, which speeds up the detection process~\cite{Samuel2019Learning}. The effect of batch size on run time can be seen in Table~\ref{table_1}. The results in Fig.~\ref{fig_8} show that the proposed OBMNet and the SVM-based method outperform the Bussgang-based linear receivers. At high SNRs, the BER floor of OBMNet detector is slightly lower than that of the SVM-based method. For the case of QPSK, $K=4$, and $N=32$, OBMNet has $10$ layers ($L=10$) with the following trained step sizes:
\begin{align*}
	&\alpha_1 = 0.32309037,\ \alpha_2 = 0.73965085,\ \alpha_3 = 0.24251865,\\
	&\alpha_4 = 0.30109185,\ \alpha_5 = 0.16300564,\ \alpha_6 = 0.11734936,\\
	&\alpha_7 = 0.09769627,\ \alpha_8 = 1.74219070,\ \alpha_9 = 0.17543483,\\
	&\alpha_{10} = 0.07491712.
\end{align*}
For the case of 16-QAM, $K=8$, and $N=128$, OBMNet has $15$ layers ($L=15$) with the following trained step sizes:
\begin{align*}
	&\alpha_1 = 0.67756593,\ \alpha_2 = 1.35809150,\ \alpha_3 = 0.83908420,\\
	&\alpha_4 = 1.16670950,\ \alpha_5 = 1.02385840,\ \alpha_6 = 1.37275460,\\
	&\alpha_7 = 0.60130936,\ \alpha_8 = 0.98949670,\ \alpha_9 = 1.25742690,\\
	&\alpha_{10} = 0.67903227,\ \alpha_{11} = 1.15905560,\ \alpha_{12} = 0.60137373,\\
	&\alpha_{13} = 0.73523980,\ \alpha_{14} = 0.33911410,\ \alpha_{15} = 0.14425066.
\end{align*}
\begin{table}[t!]
	\centering
	\caption{First stage average run time.\label{table_1}}
	\renewcommand{\arraystretch}{1.4}
	\begin{tabular}{|c|p{1.27cm} p{1.27cm} p{1.4cm} p{1.95cm}|}
		\hline
		\multicolumn{5}{|c|}{\cellcolor{cellColor}\textbf{QPSK}, $K = 4$, $N=32$} \\
		\hline
		\textbf{\makecell{batch\\size}}& \textbf{\makecell{proposed\\BZF}} & \textbf{\makecell{proposed\\BMMSE}} & \textbf{\makecell{proposed\\DNN-based}} & \textbf{\makecell{SVM-based\\\cite{Nguyen2020SVM}}} \\
		\hline
		$1$ & $1.3\times10^{-5}$ & $1.5\times10^{-5}$ & $2.2\times10^{-4}$ & $[3.1, 3.8]\times10^{-4}$ \\ \hline
		$10$& $1.1\times10^{-5}$ & $1.1\times10^{-5}$ & $5.8\times10^{-5}$ & $[3.1, 3.8]\times10^{-4}$ \\ \hline
		$100$& $1.0\times10^{-5}$ & $1.0\times10^{-5}$ & $4.2\times10^{-5}$ & $[3.1, 3.8]\times10^{-4}$ \\ \hline
		$250$& $1.0\times10^{-5}$ & $1.0\times10^{-5}$ & $3.6\times10^{-5}$ & $[3.1, 3.8]\times10^{-4}$ \\ \hline
		\hline
		\multicolumn{5}{|c|}{\cellcolor{cellColor}\textbf{16-QAM}, $K = 8$, $N=128$} \\
		\hline
		\textbf{\makecell{batch\\size}}& \textbf{\makecell{proposed\\BZF}} & \textbf{\makecell{proposed\\BMMSE}} & \textbf{\makecell{proposed\\DNN-based}} & \textbf{\makecell{SVM-based\\\cite{Nguyen2020SVM}}} \\
		\hline
		$1$ & $2.8\times10^{-5}$ & $3.5\times10^{-5}$ & $5.2\times10^{-4}$ & $[6.4,9.6]\times10^{-4}$ \\ \hline
		$5$& $2.5\times10^{-5}$ & $3.3\times10^{-5}$ & $3.1\times10^{-4}$ & $[6.4,9.6]\times10^{-4}$ \\ \hline
		$10$ & $2.4\times10^{-5}$ & $3.2\times10^{-5}$ & $2.8\times10^{-4}$ & $[6.4,9.6]\times10^{-4}$ \\ \hline
		$25$& $2.4\times10^{-5}$ & $3.2\times10^{-5}$ & $2.6\times10^{-4}$ & $[6.4,9.6]\times10^{-4}$ \\ \hline
	\end{tabular}
\end{table}
\begin{figure*}[t!]
	\centering
	\begin{subfigure}[t]{0.33\textwidth}
		\centering
		\includegraphics[width=\linewidth]{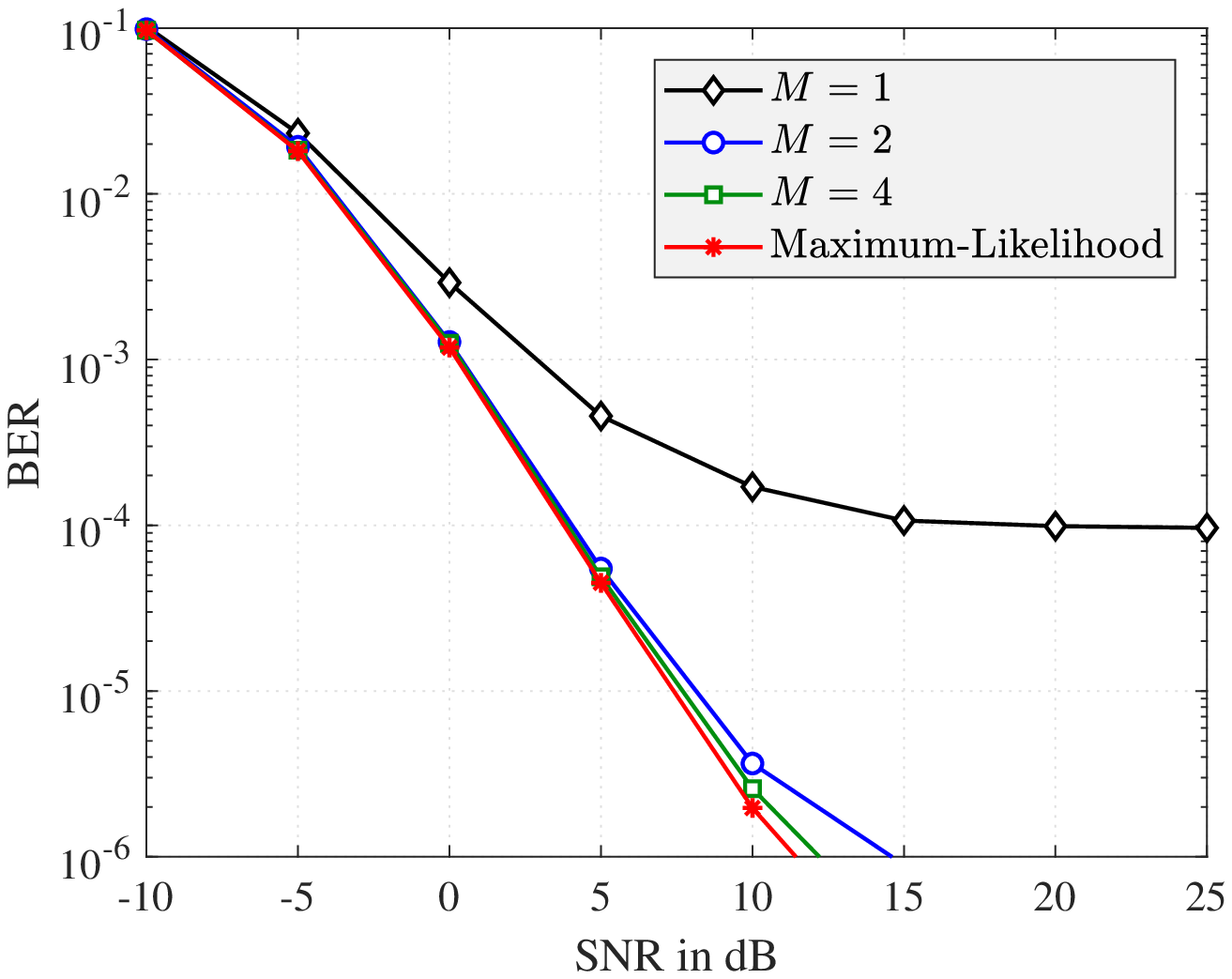}
		\caption{Proposed BZF.}
		\label{fig_9a}
	\end{subfigure}%
	\begin{subfigure}[t]{0.33\textwidth}
		\centering
		\includegraphics[width=\linewidth]{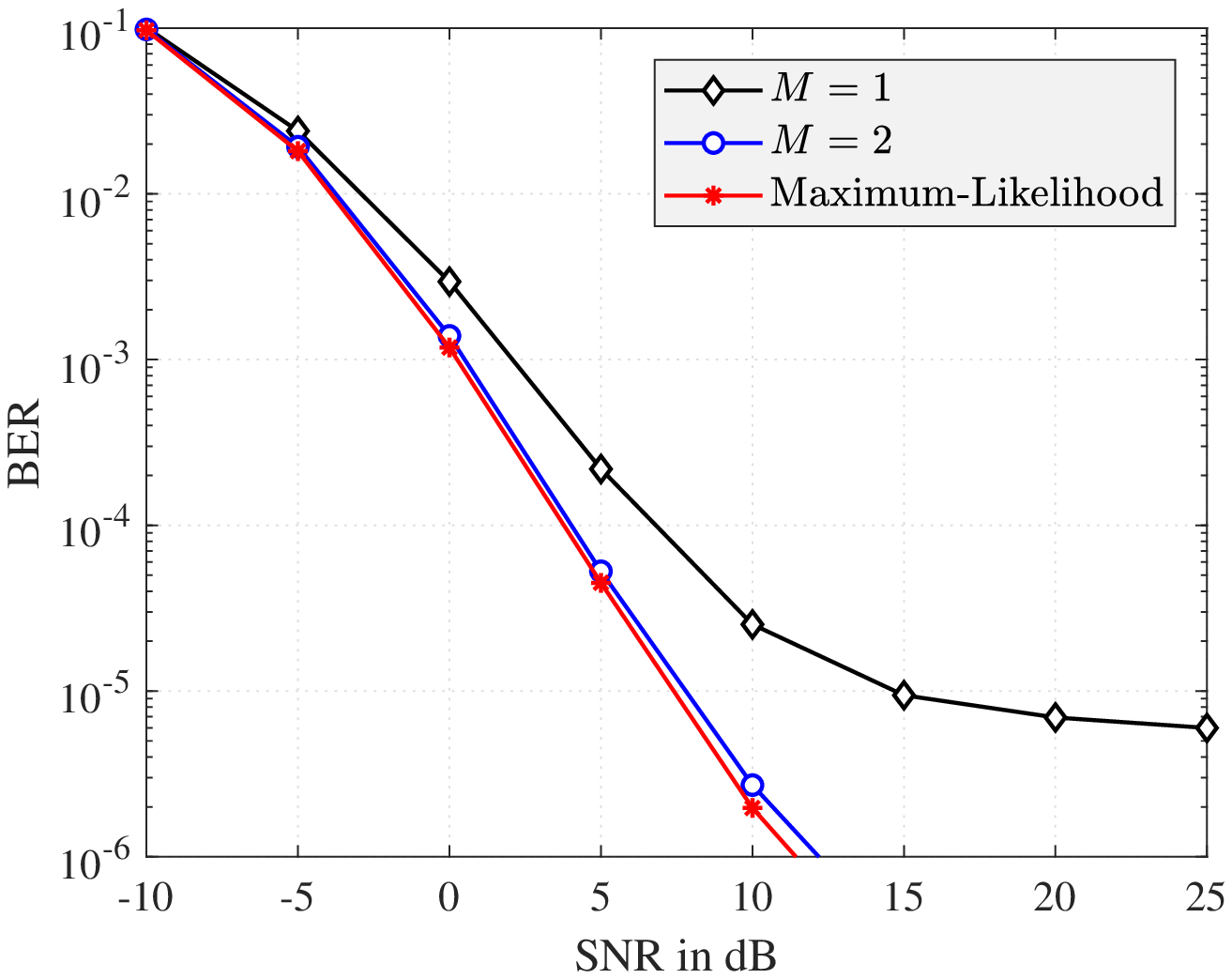}
		\caption{Proposed DNN-based.}
		\label{fig_9b}
	\end{subfigure}
	\begin{subfigure}[t]{0.33\textwidth}
		\centering
		\includegraphics[width=\linewidth]{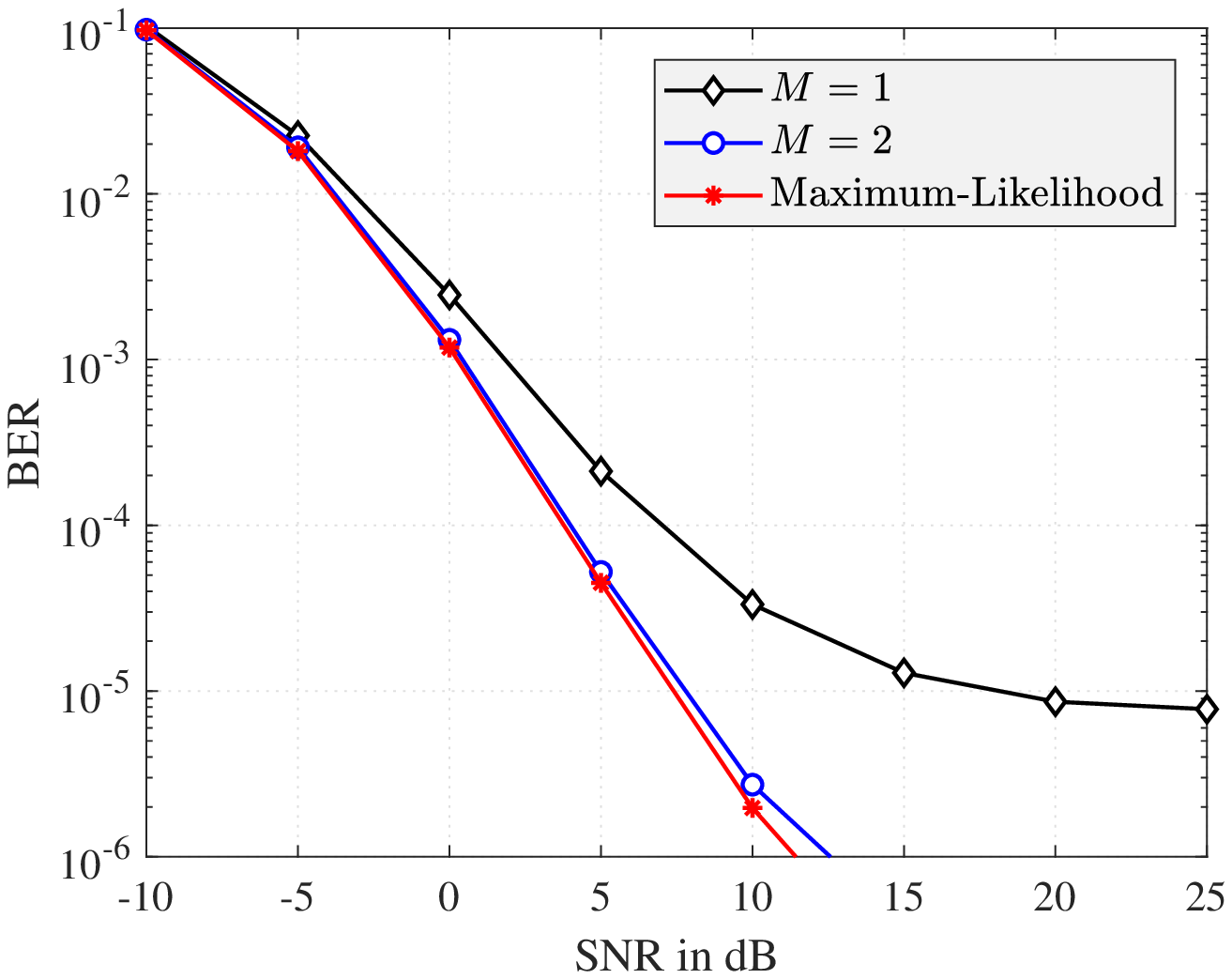}
		\caption{SVM-based~\cite{Nguyen2020SVM}.}
		\label{fig_9c}
	\end{subfigure}
	\caption{Second stage performance comparison between different receivers with $K=4$, $N=32$, and QPSK signaling.}
	\label{fig_9}
\end{figure*}
\begin{figure*}[t!]
	\centering
	\begin{subfigure}[t]{0.33\textwidth}
		\centering
		\includegraphics[width=\linewidth]{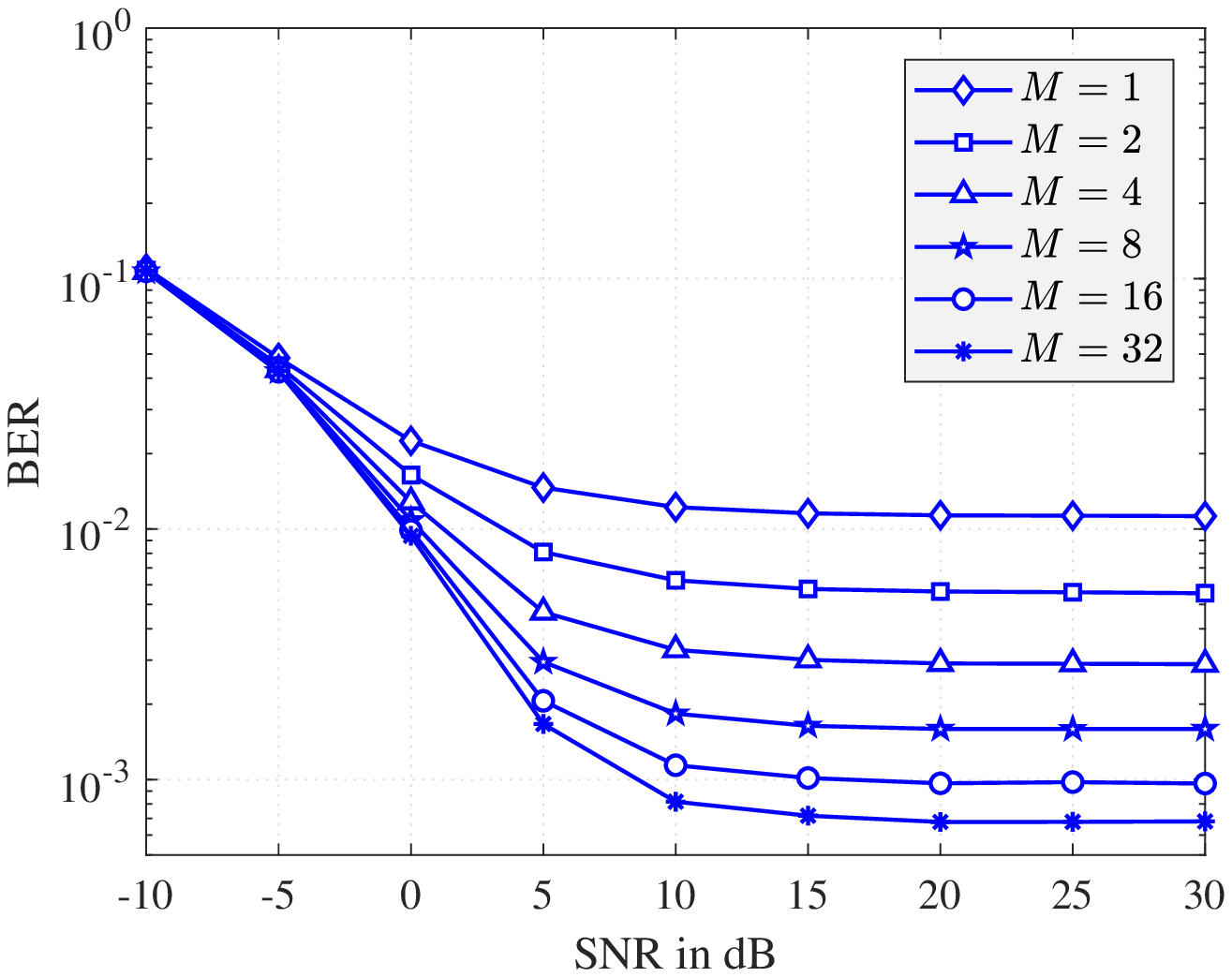}
		\caption{Proposed BZF.}
		\label{fig_10a}
	\end{subfigure}%
	\begin{subfigure}[t]{0.33\textwidth}
		\centering
		\includegraphics[width=\linewidth]{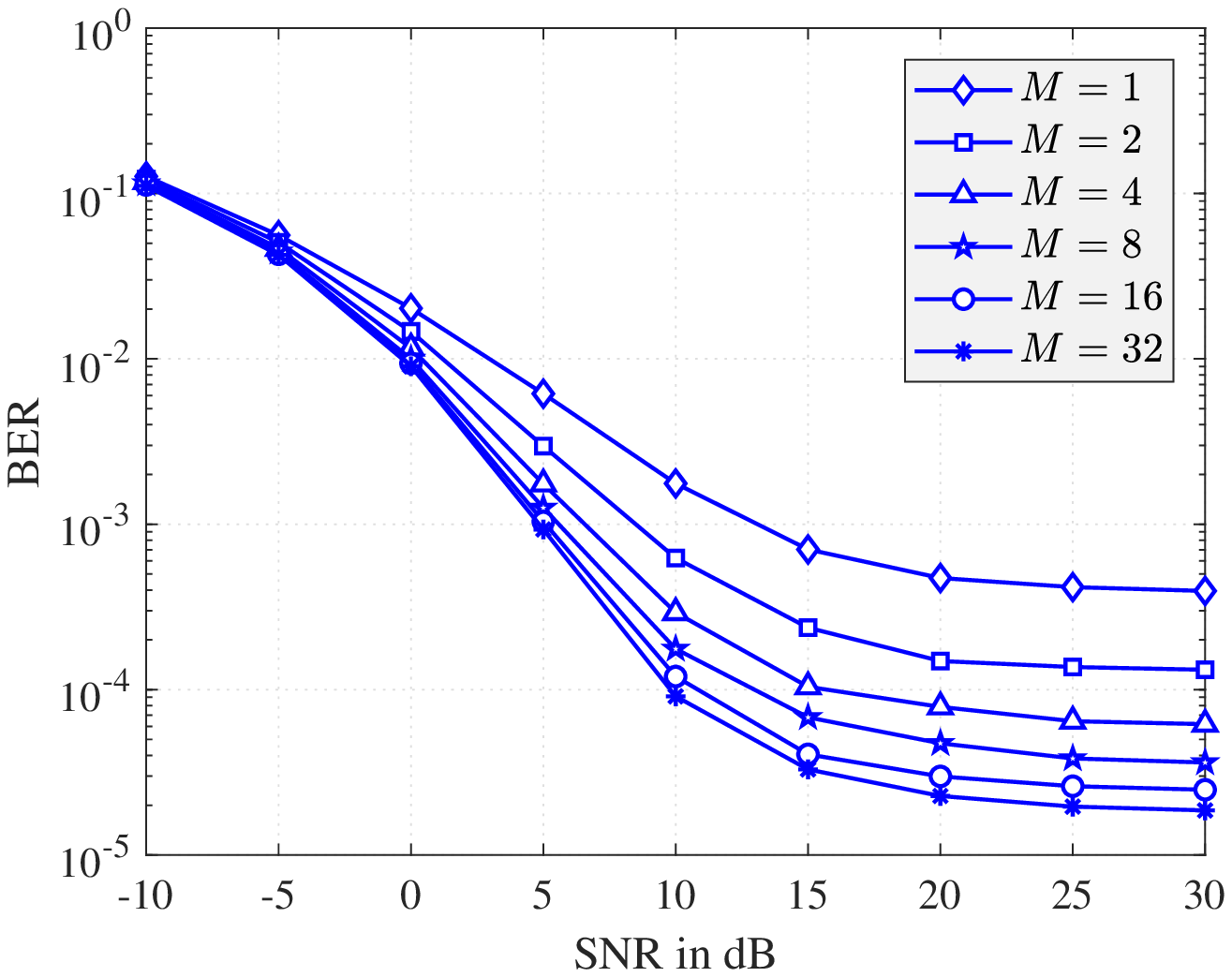}
		\caption{Proposed DNN-based.}
		\label{fig_10b}
	\end{subfigure}
	\begin{subfigure}[t]{0.33\textwidth}
		\centering
		\includegraphics[width=\linewidth]{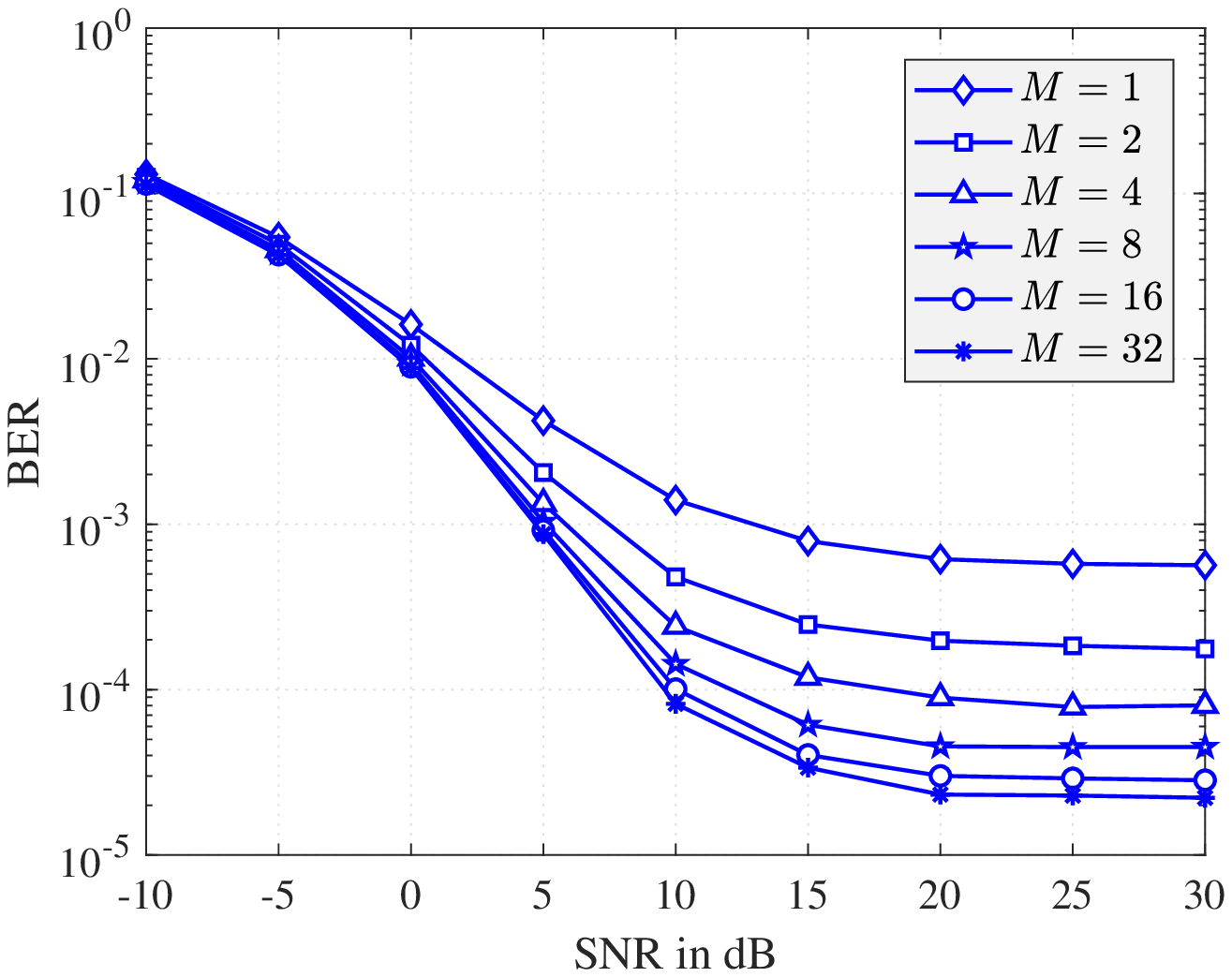}
		\caption{SVM-based~\cite{Nguyen2020SVM}.}
		\label{fig_10c}
	\end{subfigure}
	\caption{Second stage performance comparison between different receivers with $K=8$, $N=128$, and 16-QAM signaling.}
	\label{fig_10}
\end{figure*}

To evaluate the computational complexity of the receivers used in Fig.~\ref{fig_8}, average run time is reported in Table~\ref{table_1}. Since the run time is largely affected by implementation details and the associated hardware/platform, to ensure fairness, we implemented all the receivers using the same simulation hardware with Python 3.7 and the Numpy package. Note that the run time of the SVM-based method depends on the SNR, and so we report the resulting range of run times. It can be seen from Table~\ref{table_1} that the Bussgang-based linear receivers have lower complexity than OBMNet and the SVM-based receiver. This is obvious since the linear receivers only require a matrix-vector multiplication for detecting each received signal. The run time of the BZF receiver is smaller than that of BMMSE because the combining matrix $\mathbf{W}_\mathtt{BZF}$ only involves the inversion of a $K\times K$ matrix while $\mathbf{W}_\mathtt{BMMSE}$ requires the inverse of an $N\times N$ matrix. OBMNet is more computationally expensive than the linear receivers but its complexity is still much less than that of the SVM-based method. It can also be seen that the run time of OBMNet can be significantly reduced by increasing the batch size. A similar observation about the effect of the batch size on run time is reported in~\cite{Samuel2019Learning}. Note that the run time of the SVM-based method does not depend on the batch size since it processes different received signals separately and each time slot requires the SVM-based method to solve a new optimization problem. 

For the second stage, performance comparisons are given in Fig.~\ref{fig_9} for the case of QPSK with $K=4$ and $N=32$, and Fig.~\ref{fig_10} for the case of 16-QAM with $K=8$ and $N=128$.
We set $\gamma = \frac{1}{2\sqrt{2}}$ for QPSK and $\gamma = \frac{1}{2\sqrt{10}}$ for 16-QAM. Here, we compare the BZF, OBMNet, and SVM-based receivers and omit BMMSE since the performance of BZF and BMMSE are comparable, and the complexity of BZF is lower than that of BMMSE. The case of $M=1$ is equivalent to the use of symbol-by-symbol detection in the first stage. In this case, OBMNet provides the best performance, i.e., it yields the best initial detection results. When increasing $M$, the proposed NN search method in the second stage significantly improves the performance compared to the first stage. In Fig.~\ref{fig_9}, the BERs obtained with a small $M$, e.g., $M=2$, are already close to the BER of the ML detection approach. The results in Fig.~\ref{fig_10} clearly show that the performance can be improved by increasing $M$, but this requires more computation resources as seen in Table~\ref{table_2}. Thus, one should choose $M$ to balance the detection accuracy and computational complexity. It should be noted that $|\mathcal{A}|$ is always a power of two, but $M$ can be any positive integer number.
\begin{table}[t!]
	\centering
	\caption{Second stage average run time.\label{table_2}}
	\renewcommand{\arraystretch}{1.4}
	\begin{tabular}{|c|p{2.1cm} p{2.1cm} p{2.3cm}|}
		\hline
		\multicolumn{4}{|c|}{\cellcolor{cellColor}\textbf{QPSK}, $K = 4$, $N=32$, $\text{batch size} = 250$} \\
		\hline
		$M$ & \textbf{\makecell{proposed\\BZF}} & \textbf{\makecell{proposed\\DNN-based}} & \textbf{\makecell{SVM-based\\\cite{Nguyen2020SVM}}} \\
		\hline
		$2$ & $[0.5, 1.0]\times10^{-4}$ & $[0.4, 1.0]\times10^{-4}$ & $[0.6, 1.2]\times10^{-4}$ \\ \hline
		\hline
		\multicolumn{4}{|c|}{\cellcolor{cellColor}\textbf{16-QAM}, $K = 8$, $N=128$, $\text{batch size} = 25$} \\
		\hline
		$M$& \textbf{\makecell{proposed\\BZF}} & \textbf{\makecell{proposed\\DNN-based}} & \textbf{\makecell{SVM-based\\\cite{Nguyen2020SVM}}} \\
		\hline
		$2$ & $[2.0, 2.5]\times10^{-4}$ & $[1.6, 2.5]\times10^{-4}$ & $[1.9,3.2]\times10^{-4}$ \\ \hline
		$4$ & $[2.8, 3.5]\times10^{-4}$ & $[1.8, 3.7]\times10^{-4}$ & $[2.1,5.2]\times10^{-4}$ \\ \hline
		$8$ & $[3.9, 6.2]\times10^{-4}$ & $[2.0, 6.6]\times10^{-4}$ & $[2.4,9.6]\times10^{-4}$ \\ \hline
		$16$& $[5.4, 13.1]\times10^{-4}$ & $[2.3, 14.7]\times10^{-4}$ & $[3.3,21.7]\times10^{-4}$ \\ \hline
		$32$& $[8.1, 30.0]\times10^{-4}$ & $[3.0, 34.1]\times10^{-4}$ & $[4.3,46.5]\times10^{-4}$ \\ \hline
	\end{tabular}
\end{table}
\section{Conclusion}
\label{sec_conclusion}
In this paper, we have proposed two-stage detection methods for massive MIMO systems with one-bit ADCs. In particular, for the first stage, we proposed new linear receivers based on the Bussgang decomposition and a novel model-driven OBMNet detector, which is constructed based on a reformulated robust ML detection problem. The layered structure of OBMNet is simple, unique, and adaptive to the CSI and received signals. These receivers outperform existing approaches and also have low complexity. For the second stage, an NN search method was proposed to further improve the performance of the first stage. This NN search method allows one to limit the search complexity as desired.

\appendices
\section{Proof of Proposition~\ref{proposition1}}
\label{append1}
Since $\mathbf{x}_m$ is the $m^\mathrm{th}$ nearest symbol vector, we have the following condition:
\begin{equation}
\|\mathbf{x}_{1}-\tilde{\mathbf{x}}\|^2 < \ldots <\|\mathbf{x}_{m-1}-\tilde{\mathbf{x}}\|^2 < \|\mathbf{x}_m-\tilde{\mathbf{x}}\|^2 < \|\mathbf{x}-\tilde{\mathbf{x}}\|^2
\label{eq_mth_nearest_condition}
\end{equation}
for any $\mathbf{x}\notin \mathcal{X}_m$.

We prove the proposition by contradiction. Suppose that $\mathbf{x}_m$ is not a neighbor of $\mathcal{X}_{m-1}$, i.e., $\mathbf{x}_m \notin \mathcal{N}(\mathcal{X}_{m-1})$ or $d_\mathrm{min}(\mathbf{x}_m, \mathcal{X}_{m-1})>1$. For the sake of simplicity, we consider the case where $d_\mathrm{min}(\mathbf{x}_m, \mathcal{X}_{m-1}) = 2$. Proof for the other cases where $d_\mathrm{min}(\mathbf{x}_m, \mathcal{X}_{m-1}) > 2$ can be accomplished similarly. 

Let $\mathbf{x}_{p}\in \mathcal{X}_{m-1}$ with $p\in\{1,2,\ldots,m-1\}$ be a symbol vector such that $d(\mathbf{x}_p,\mathbf{x}_m)=2$. Without loss of generality, we can always assume that the two position indices at which the differences occur are~$1$ and~$2$, i.e.,
\begin{equation}
\begin{cases}
x_{m,1} \neq x_{p,1}\\
x_{m,2} \neq x_{p,2}\\
x_{m,i} = x_{p,i} \ \forall i \in \{3,\ldots,2K\}.
\end{cases}
\end{equation}
Now, we consider two other symbol vectors $\mathbf{x}' = [x'_1,\ldots,x'_{2K}]^T$ and $\mathbf{x}'' = [x''_1,\ldots,x''_{2K}]^T$ such that
\begin{equation}
\begin{cases}
x'_1 = x_{m,1} \neq x_{p,1}=x''_1\\
x'_2 = x_{p,2} \neq x_{m,2}=x''_2\\
x'_i = x''_i = x_{p,i} = x_{m,i} \ \forall i \in \{3,\ldots,2K\}.
\end{cases}
\label{eq_xprime_and_xdoubleprime}
\end{equation}
Hence, $\mathbf{x}'$ and $\mathbf{x}''$ are the two symbol vectors satisfying $d(\mathbf{x}',\mathbf{x}_m) = d(\mathbf{x}'',\mathbf{x}_m) = 1$. In other words, both $\mathbf{x}'$ and $\mathbf{x}''$ are neighbors of $\mathbf{x}_m$.

If $\mathbf{x}'\in \mathcal{X}_{m-1}$ and/or $\mathbf{x}''\in \mathcal{X}_{m-1}$, then $d_\mathrm{min}(\mathbf{x}_m, \mathcal{X}_{m-1}) = 1$ because $\mathbf{x}_m$ is a neighbor of both $\mathbf{x}'$ and $\mathbf{x}''$, which is contradicted by the assumption that $d_\mathrm{min}(\mathbf{x}_m, \mathcal{X}_{m-1}) = 2$. Thus, $\mathbf{x}_m$ is a neighbor of $\mathcal{X}_{m-1}$, i.e, $\mathbf{x}_m \in \mathcal{N}(\mathcal{X}_{m-1})$. 

If $\mathbf{x}'\notin \mathcal{X}_{m-1}$ and $\mathbf{x}''\notin \mathcal{X}_{m-1}$, we have 
\begin{equation}
|x_{m,1}-\tilde{x}_1|^2 = |x'_1-\tilde{x}_1|^2 > |x_{p,1}-\tilde{x}_1|^2.
\label{eq_xprime_condition}
\end{equation}
Adding both sides of~\eqref{eq_xprime_condition} with $|x_{m,2}-\tilde{x}_2|^2$ yields
\begin{equation*}
|x_{m,1}-\tilde{x}_1|^2 + |x_{m,2}-\tilde{x}_2|^2 > |x_{p,1}-\tilde{x}_1|^2 + |x_{m,2}-\tilde{x}_2|^2,
\end{equation*}
which can be rewritten as
\begin{equation}
|x_{m,1}-\tilde{x}_1|^2 + |x_{m,2}-\tilde{x}_2|^2 > |x''_1-\tilde{x}_1|^2 + |x''_2-\tilde{x}_2|^2
\label{eq_xm1_vs_xdoubleprime_condition}
\end{equation}
because $x_{p,1}=x''_1$ and $x_{m,2}=x''_2$. The inequality in~\eqref{eq_xm1_vs_xdoubleprime_condition} indicates that $\|\mathbf{x}_m-\tilde{\mathbf{x}}\|^2 > \|\mathbf{x}''-\tilde{\mathbf{x}}\|^2$, which means $\mathbf{x}''$ is closer to $\tilde{\mathbf{x}}$ than $\mathbf{x}_m$, or in other words, $\mathbf{x}_m$ is not the $m^\mathrm{th}$ nearest symbol vector of $\tilde{\mathbf{x}}$. This is contradicted by~\eqref{eq_mth_nearest_condition}.

\bibliographystyle{IEEEtran}
\bibliography{ref}

\end{document}